\documentclass[12pt,preprint]{aastex}
\usepackage{lscape,graphicx,rotating}

\newcommand{\Swift}{\textit{Swift}}

\newcommand{\hst}{\textit{HST}}

\newcommand{\Rc}{\textit{R$_{\mathrm{C}}$}}

\newcommand{\Vc}{\textit{V$_{\mathrm{C}}$}}

\newcommand{\ip}{\textit{i$^{\prime}$}}
\newcommand{\zp}{\textit{z$^{\prime}$}}
\newcommand{\gp}{\textit{g$^{\prime}$}}
\newcommand{\rp}{\textit{r$^{\prime}$}}
\newcommand{\J}{\textit{J}}
\newcommand{\Hm}{\textit{H}}
\newcommand{\Ks}{\textit{K$_{s}$}}
\newcommand{\sci}{Science}
\newcommand{\bai}{Bulletins of the Astronomical Society of India}

\begin{document}

\title{The Collimation and Energetics of the Brightest \Swift\ Gamma-Ray
Bursts}

\author{S.~B.~Cenko\altaffilmark{1}, D.~A.~Frail\altaffilmark{2},
  F.~A.~Harrison\altaffilmark{3}, S.~R.~Kulkarni\altaffilmark{4},
  E.~Nakar\altaffilmark{5}, P.~Chandra\altaffilmark{6},
  N.~R.~Butler\altaffilmark{1}, D.~B.~Fox\altaffilmark{7},
  A.~Gal-Yam\altaffilmark{8}, M.~M.~Kasliwal\altaffilmark{4},
  J.~Kelemen\altaffilmark{9}, D.-S.~Moon\altaffilmark{10},
  P.~A.~Price\altaffilmark{11}, A.~Rau\altaffilmark{12,4},
  A.~M.~Soderberg\altaffilmark{13,14}, H.~I.~Teplitz\altaffilmark{15},
  M.~W.~Werner\altaffilmark{15}, D.~C.-J.~Bock\altaffilmark{16},
  J.~S.~Bloom\altaffilmark{1}, D.~A.~Starr\altaffilmark{1,17},
  A.~V.~Filippenko\altaffilmark{1}, 
  R.~A.~Chevalier\altaffilmark{18}, N.~Gehrels\altaffilmark{19},
  J.~N.~Nousek\altaffilmark{20}, and T.~Piran\altaffilmark{21}}

\altaffiltext{1}{Department of Astronomy, 601 Campbell Hall, University of
  California, Berkeley, CA 94720-3411, USA.}
\altaffiltext{2}{National Radio Astronomy Observatory, P.O.~Box 0, 1003
  Lopezville Road, Socorro, NM 87801, USA.}
\altaffiltext{3}{Space Radiation Laboratory, California Institute of 
  Technology, M/C 249-17, Pasadena, CA 91125, USA.}
\altaffiltext{4}{Department of Astronomy, California Institute of Technology,
  M/C 249-17, Pasadena, CA 91125, USA.}
\altaffiltext{5}{Raymond and Beverly Sackler School of Physics \& Astronomy, 
  Tel Aviv University, Tel Aviv 69978, Israel.}
\altaffiltext{6}{Department of Physics, Royal Military College of Canada,
  Kingston, ON, Canada.}
\altaffiltext{7}{Department of Astronomy \& Astrophysics, 525 Davey
  Laboratory, Pennsylvania State University, University Park, PA 16802, USA.}
\altaffiltext{8}{Benoziyo Center for Astrophysics, Weizmann Institute of 
  Science, 76100 Rehovot, Israel.}
\altaffiltext{9}{Konkoly Observatory, H-1525, Box 67, Budapest, Hungary.}
\altaffiltext{10}{Department of Astronomy and Astrophysics, University of 
  Toronto, Toronto, ON M5S 3H4, Canada.}
\altaffiltext{11}{Institute for Astronomy, University of Hawaii, 2680 
  Woodlawn Drive, Honolulu, HI 96822, USA.}
\altaffiltext{12}{Max-Planck Institute for Extra-Terrestrial Physics,
  Giessenbachstr.~1, 85748 Garching, Germany.}
\altaffiltext{13}{Harvard-Smithsonian Center for Astrophysics, 60 Garden 
  Street, Cambridge, MA 02138, USA.}
\altaffiltext{14}{Hubble Fellow.}
\altaffiltext{15}{Spitzer Science Center, California Institute of Technology, 
  Pasadena, CA 91125, USA.}
\altaffiltext{16}{Combined Array for Research in Millimeter-wave Astronomy, 
  P.O.~Box 968, Big Pine, CA 93513, USA.}
\altaffiltext{17}{Las Cumbres Observatory Global Telescope Network, Inc., 
  6740 Corona Dr.~Suite 102, Santa Barbara, CA 93117, USA.}
\altaffiltext{18}{Department of Astronomy, University of Virgina, P.O.~Box
  400325, Charlottesville, VA 22904, USA.}
\altaffiltext{19}{Astrophysics Science Division, Code 660.1, NASA/Goddard 
  Space Flight Center, Greenbelt, MD 20770, USA.}
\altaffiltext{20}{Department of Astronomy \& Astrophysics, Pennsylvania 
  State University, 104 Davey Laboratory, University Park, PA 16802, USA.}
\altaffiltext{21}{The Racah Institute of Physics, Hebrew University, 
  Jerusalem 91904, Israel.}

\slugcomment{Submitted to \textit{ApJ}}

\shorttitle{The Brightest \Swift\ GRBs}
\shortauthors{Cenko \textit{et al.}}

\begin{abstract}
Long-duration $\gamma$-ray bursts (GRBs) are widely believed to be 
highly-collimated explosions (opening angle $\theta \approx$ 1--10$^{\circ}$).  
As a result of this beaming factor, the true energy release from a GRB is 
usually several orders of magnitude smaller than the observed isotropic value.  
Measuring this opening angle, typically inferred from an achromatic steepening
in the afterglow light curve (a ``jet'' break), has proven exceedingly
difficult in the \Swift\ era.  Here we undertake a study of five of the 
brightest (in terms of the isotropic prompt $\gamma$-ray energy release,
$E_{\gamma,\mathrm{iso}}$) GRBs in the \Swift\ era to search for jet breaks and 
hence constrain the collimation-corrected energy release.  We present 
multi-wavelength (radio through X-ray) observations of GRBs\,050820A, 060418, and 
080319B, and construct afterglow models to extract the opening angle and 
beaming-corrected energy release for all three events.  Together with results 
from previous analyses of GRBs\,050904 and 070125, we find evidence for an 
achromatic jet break in all five events, strongly supporting the canonical 
picture of GRBs as collimated explosions.  The most natural explanation for
the lack of observed jet breaks from most \Swift\ GRBs is therefore selection 
effects.  However, the opening angles for the events in our sample are larger 
than would be expected if all GRBs had a canonical energy release of 
$\sim 10^{51}$\,erg.  The total energy release we measure for those 
``hyper-energetic'' ($E_{\mathrm{tot}} \gtrsim 10^{52}$\,erg) events in our 
sample is large enough to start challenging models with a magnetar as the 
compact central remnant.
\end{abstract}

\keywords{gamma-rays: bursts --- X-rays: individual (GRB\,050820A; GRB\,050904;
          GRB\,060418; GRB\,070125; GRB\,080319B)}

\section{Introduction}
\label{sec:intro}
Accurate calorimetry is fundamental to understanding any astrophysical
phenomenon.  In the case of long-duration $\gamma$-ray bursts 
(GRBs)\footnote{Throughout this work, we use the term ``long-duration'' GRBs 
to refer to those events that arise from the core collapse of a massive star
\citep{w93,wb06}, despite the fact that duration alone is not sufficient to
distinguish from those GRBs associated with an older stellar population
(e.g., \citealt{dls+06,zzv+09}).}, three measurements are required for an
accounting of the total relativistic\footnote{We neglect contributions from
slower moving material (i.e., supernova emission) as well as 
non-electromagnetic emission (neutrinos, gravitational radiation, etc.).}
energy release: (1) $E_{\gamma,\mathrm{iso}}$, the isotropic energy release in the
prompt $\gamma$-ray emission, which is inferred from the $\gamma$-ray fluence
measured by the detecting satellite and the associated afterglow or host
redshift; (2) $\theta$, the half-opening angle of the bipolar conical outflow, which is 
inferred from the detection of a characteristic achromatic steepening in the afterglow 
light curve (i.e., a ``jet'' break; \citealt{r99,sph99}); and, (3) 
$E_{\mathrm{KE}}$, the kinetic energy of the shock powering the afterglow 
emission, which can be inferred either via broadband afterglow modeling 
(e.g., \citealt{pk02,yhs+03}), or, more accurately, from late-time radio 
calorimetry in the non-relativistic phase (e.g., \citealt{bkf04,fsk+05}).

Compilations of such measurements for the first GRB afterglows suggested that the
collimation-corrected energy release, either from the prompt $\gamma$-rays 
($E_{\gamma}$) or powering the afterglow ($E_{\mathrm{KE}}$), was tightly
clustered around $\sim 10^{51}$\,erg \citep{fks+01,bkf03,bfk03}.  This result
helped to establish the connection between GRBs and massive stars, as 
core-collapse supernovae (SNe) result in a comparable output of kinetic
energy.  It further motivated efforts to utilize GRBs as standardizable 
candles to constrain the cosmological model of our universe (e.g., 
\citealt{dlx04,fag+06,s07}), much as has been done for Type Ia SNe 
(\citealt{rfc+98,pag+99}; see, e.g., \citealt{f05} for a review).

It was soon realized, however, that the most nearby (redshift $z \lesssim 0.1$) GRBs
were several orders of magnitude less energetic than the typical GRB
at $z \gtrsim 1$ \citep{bfk03,skb+04}.  Furthermore, these underluminous 
events appear to be significantly more common (in terms of volumetric rate) 
than their cosmological brethren.  Though the reason for this dichotomy remains
a mystery, it suggests that perhaps long-duration GRBs are a more
diverse population than originally envisioned.

The launch of the \Swift\ satellite \citep{gcg+04} in 2004 November heralded a 
potential revolution in the study of GRB energetics.  With its unique 
combination of sensitivity ($\sim 100$ GRB localizations yr$^{-1}$, 
an order of magnitude improvement over previous satellites) and precise 
localization capabilities ($\sim 3\arcmin$ positions arrive only seconds after 
the GRB trigger, with $\sim 3\arcsec$ positions delivered minutes later), 
\Swift\ promised to deliver a tremendous increase in the number of events 
suitable for detailed studies of energetics. 

Furthermore, the onboard X-ray telescope (XRT; \citealt{bhn+05}) has provided
the first detailed look at X-ray afterglow evolution.  Before the launch
of \Swift, opening angles were typically inferred from the optical and 
occasionally radio bandpasses. X-ray afterglows, particularly at early times,
were a relatively poorly sampled phase space.  The additional leverage provided
by the X-ray regime promised to greatly simplify the task of distinguishing
jet breaks from other predicted spectral features in afterglow light curves
due to the achromatic nature of this hydrodynamical transition.

\begin{figure}[t!]
  \centerline{\plotone{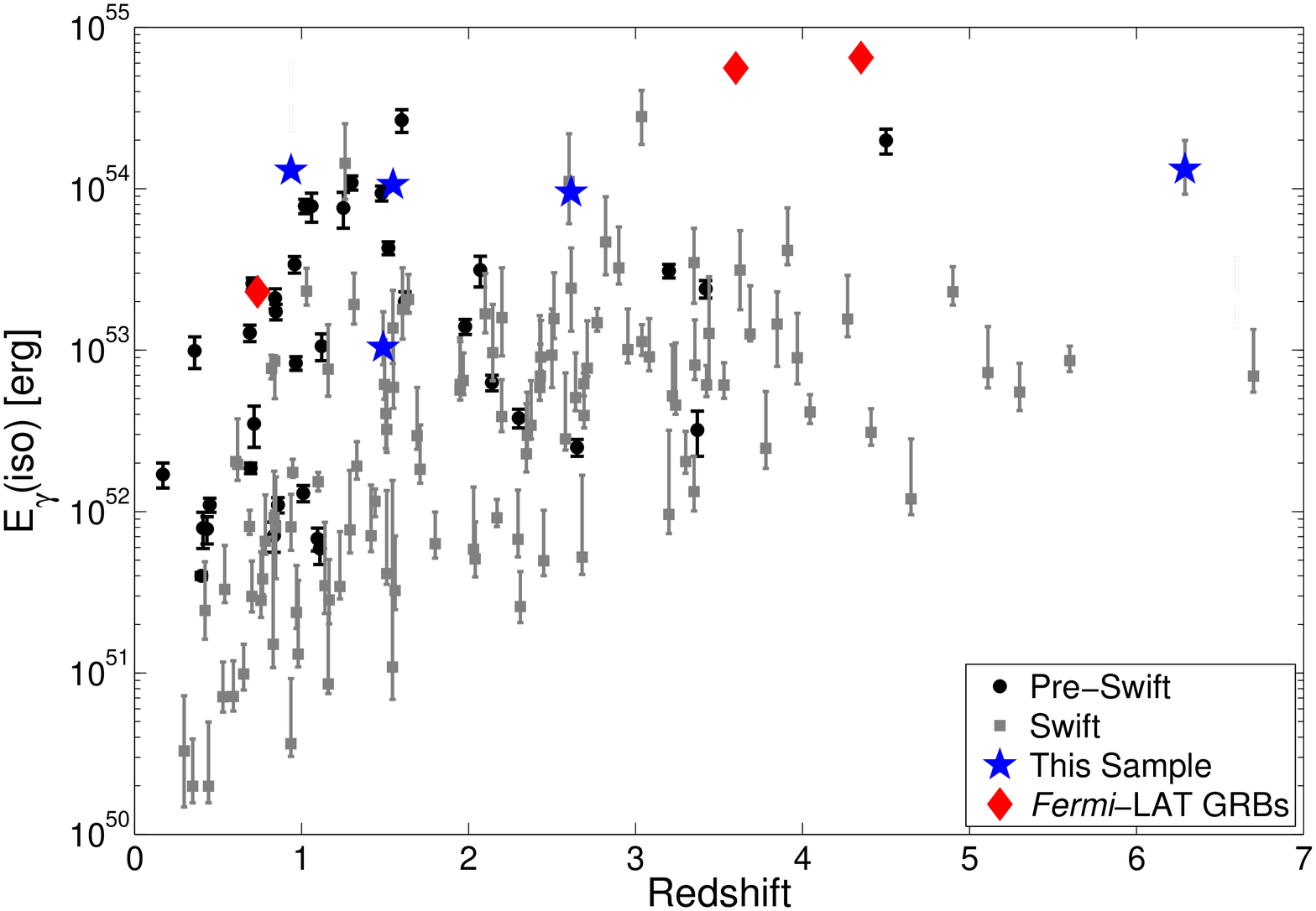}}
  \caption[Isotropic prompt $\gamma$-ray energy release 
           ($E_{\gamma,\mathrm{iso}}$) of GRBs with
           measured redshift.]
  {Isotropic prompt $\gamma$-ray energy release ($E_{\gamma,\mathrm{iso}}$) of 
   GRBs with measured redshift.  All prompt energy
   releases have been transformed to the rest-frame 1\,keV to 10\,MeV 
   bandpass.  The increased sensitivity of the \Swift\ BAT results in a 
   population with lower values of $E_{\gamma,\mathrm{iso}}$ and larger redshifts.
   It is not surprising, then, that typical \Swift\ events should have large
   (or even isotropic) opening angles, making jet break measurements
   quite difficult \citep{psf03}.  In this work we focus on those events in 
   the \Swift\ sample with the largest values of $E_{\gamma,\mathrm{iso}}$.  
   References: pre-\Swift: \citet{a06}; \Swift: \citet{bkb+07}; 
   {\it Fermi}-LAT: \citet{gck+09,GCN.9030,GCN.9057}.}
\label{fig:eiso}
\end{figure}

Despite these advances, measuring bolometric fluences of \Swift\ events has 
proven to be a challenging task.  First, the limited bandpass
(15--150\,keV) of the \Swift\ Burst Alert Telescope (BAT; \citealt{bbc+05})
captures only a fraction of the traditional $\gamma$-ray regime.  As evidenced
by Figure~\ref{fig:eiso}, the uncertainties associated with \Swift\
$E_{\gamma,\mathrm{iso}}$ measurements are significantly larger than the 
pre-\Swift\ sample, due to the difficulty in extrapolating to the traditional
1--10$^{4}$\,keV (rest frame) bolometric bandpass.  We note that the \Swift\ 
measurements shown in Figure~\ref{fig:eiso} incorporate a Bayesian prior on 
the spectral peak energy ($E_{\mathrm{p}}$) based on the $E_{\mathrm{p}}$ 
distribution measured by the BATSE instrument (see \citealt{bkb+07} for 
details).  Without this constraint the $E_{\gamma,\mathrm{iso}}$ measurements 
would be even more uncertain.
 
Second, the detailed X-ray light curves provided by the \Swift\ XRT have
revealed a central engine capable of injecting energy into the forward 
shock at late times ($t \gg \Delta t_{\mathrm{GRB}}$), either as short-lived 
X-ray flares that can contain a comparable amount of energy to $E_{\gamma}$ 
\citep{brf+05}, or as extended periods of shallow decay (so-called ``plateau'' 
phases) inconsistent with standard afterglow models 
\citep{fp06,nkg+06,zfd+06}.  While alternative interpretations for both 
phenomena exist, these discoveries suggest that our simplistic adiabatic 
picture of afterglow evolution may need to be revised.

Most importantly, surprisingly few \Swift\ afterglows have shown the 
characteristic achromatic steepening associated with a collimated outflow.
Several groups have conducted a comprehensive analysis of a large
sample of X-ray \citep{p07,kb08,rlb+08} and/or optical \citep{lrz+08} light
curves, finding that at most only a small fraction exhibit clear evidence for
collimation.   Without these collimation corrections, the true energy release 
from \Swift\ events has remained highly uncertain (e.g., 
\citealt{kb08,rlb+08}).

Here we take a different approach.  To begin with, we focus only on those 
\Swift\ events with the largest values of $E_{\gamma,\mathrm{iso}}$
(Fig.~\ref{fig:eiso}).  In the framework of a canonical GRB energy release,
these events should have the smallest opening angles, thereby easing to some
extent the observational bias against late-time jet breaks.  Alternatively, if
isotropic, these extreme events would place the strongest constraints on the 
mechanism powering these explosions.  Such high-fluence events are also more 
likely to be detected by other $\gamma$-ray satellites, providing additional 
coverage in the traditional $\gamma$-ray bandpass and thereby better 
constraining the prompt $\gamma$-ray energy release.

In addition, we only consider GRB afterglows with broadband (X-ray, optical,
{\it and} radio) coverage extending out to late times ($t \gtrsim 1$\,month).
The radio bandpass is particularly sensitive to wide-angle jets, as the 
synchrotron peak frequency typically does not reach the radio bandpass until
days or even weeks after the burst, when the X-ray and optical bands may be
too faint to detect a jet break.  Well-sampled, broadband light curves ensure
accurate constraints on both the opening angle and the kinetic energy
powering the afterglow.

Given these constraints, we include five events in our \Swift\ sample:
GRBs 050820A, 050904, 060418, 070125, and 080319B.  This sample is not meant 
to be representative of the \Swift\ population as a whole.  Nor, for that
matter, have we included {\it all} of the \Swift\ events with large 
$E_{\gamma,\mathrm{iso}}$ values, as most lack the radio and late-time
optical coverage necessary for afterglow modeling (e.g., GRB\,061007; 
\citealt{sdp+07b}).  Instead, we argue that great insight, in particular with 
regard to progenitor models, can come from studies of even a small number
of events at the extreme.

This work is organized as follows.  In \S~\ref{sec:obs} we present our 
observations of the afterglows of GRB\,050820A, GRB\,060418, and GRB\,080319B.  
We then construct broadband afterglow models to extract the opening angle and
afterglow energy for each one in \S~\ref{sec:models}.  To complete our 
sample, we include analogous results from previous broadband modeling of 
GRB\,050904 \citep{fck+06} and GRB\,070125 \citep{ccf+08}.  We compare the 
total energy release from these five events with the pre-\Swift\ sample
in \S~\ref{sec:energy}, and conclude with a discussion of the future of 
GRB energetics studies in \S~\ref{sec:conclusion}.   

Throughout this work, we adopt a standard $\Lambda$CDM cosmology with
$H_{0}$ = 71\,km s$^{-1}$ Mpc$^{-1}$, $\Omega_{\mathrm{m}} = 0.27$, and
$\Omega_{\Lambda} = 1 - \Omega_{\mathrm{m}} = 0.73$ \citep{sbd+07}.  We define the
flux density power-law temporal and spectral decay indices $\alpha$ and
$\beta$ as $f_{\nu} \propto t^{-\alpha} \nu^{-\beta}$ (e.g., \citealp{spn98}).
All errors quoted are 1$\sigma$ ($68\%$) confidence intervals unless
otherwise noted.

\section{Observations and Data Reduction}
\label{sec:obs}

\subsection{GRB\,050820A}
\label{sec:obs:050820A}
GRB\,050820A was remarkable in two respects.  First, the 
\Swift\ BAT triggered on a faint $\gamma$-ray precursor nearly 4\,min 
before the bulk of the prompt emission, enabling contemporaneous $\gamma$-ray, 
X-ray, and optical coverage during the GRB itself.  Both the X-ray and (to a 
lesser extent) the optical emission abruptly brightened in concert with the 
onset of the GRB, suggesting a common origin \citep{vww+06}.  

The prompt emission was 
observed by the Konus-{\it Wind} instrument, providing spectral coverage from 
20\,keV to 1\,MeV \citep{ckh+06}.  Extrapolating the observed spectrum to a 
rest-frame bandpass of 1--10$^{4}$\,MeV, we find a fluence of 
$(6.1^{+1.9}_{-0.9}) \times 10^{-5}$\,erg cm$^{-2}$.  At $z=2.615$ 
\citep{pcb+07}, the total isotropic prompt energy release in this bandpass
was $E_{\gamma,\mathrm{iso}} = (9.7^{+3.1}_{-1.4}) \times 10^{53}$\,erg.  

In addition, the X-ray and particularly optical afterglow emission from
GRB\,050820A was quite bright, allowing the decay to be traced out to late
times.  The majority of our observations of GRB\,050820A were presented by
\citet{ckh+06}.  We reported the detection of a likely jet break
at $t_{\mathrm{j}} = 18 \pm 2$\,d based on late-time {\it Hubble Space Telescope (HST)} observations,
later than nearly all previously detected jet breaks in the optical 
bandpass \citep{zkk06}.  

Here we supplement this already rich data set with additional late-time 
X-ray and optical imaging.  \Swift\ XRT data were taken from the online 
compilation of N.~Butler\footnote{http://astro.berkeley.edu/$\sim$nat/swift; see
\citet{bk07} for details.}.  These detections extend the X-ray coverage
out to $t \approx 46$\,d, well past the previously claimed jet break time.

\begin{figure}[t!]
  \centerline{\plotone{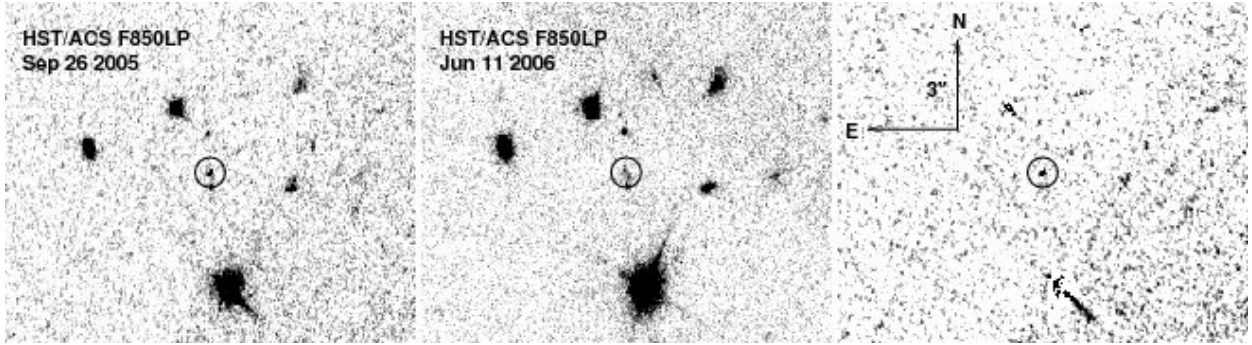}}
  \caption[\hst\ Imaging of GRB\,050820A]
  {\hst\ Imaging of GRB\,050820A.  \textit{Left:} An \textit{F850LP} ACS image
   taken on 2005 September 26.  \textit{Center:} The same field in an image
   obtained on 2006 June 11.  \textit{Right:} A digital subtraction of the
   two images, revealing the residual afterglow emission.  The host-galaxy 
   contribution to the afterglow flux at the time of the 2005 September 26 
   image is significant ($\sim 35$\%).}
\label{fig:050820Ahst}
\end{figure}

We have also obtained optical imaging of the host galaxy of GRB\,050820A
using the Wide Field Channel of the Advanced Camera for Surveys (ACS) on 
\hst\ (Fig.~\ref{fig:050820Ahst}).  Under program GO-10551 (PI: Kulkarni),
we obtained a total of 2238\,s of exposure time in the \textit{F625W} (Sloan
\rp) filter, 4404\,s of exposure time in the \textit{F775W} (Sloan \ip) 
filter, and 14280\,s of exposure time in the \textit{F850LP} (Sloan \zp)
filter beginning on 2006 June 5 (UT dates are used throughout this paper).  We processed the data using the 
\texttt{multidrizzle} routine \citep{fh02} in the \texttt{stsdas} 
IRAF \footnote{IRAF is distributed by the National Optical Astronomy 
Observatory, which is operated by the Association for Research in 
Astronomy, Inc., under cooperative agreement with the National Science 
Foundation.} package.
We used \texttt{pixfrac} $=0.8$ and \texttt{pixscale} $= 1.0$ for the 
drizzling procedure, resulting in a pixel scale of $0.05$\arcsec\ pixel$^{-1}$.
Following the recipe for 
point-source\footnote{The host galaxy is only marginally extended in the \hst\
images.} photometry from \citet{sjb+05}, we measure the following (AB)
magnitudes: $F625W = 26.04 \pm 0.13$; $F775W = 26.09 \pm 0.11$; $F850LP = 
25.91 \pm 0.11$ (including a correction for the small amount of Galactic
extinction: $E(B-V)=0.044$\,mag; \citealt{sfd98}).  These results are consistent
with, although slightly brighter than, the values reported by \citet{cpp+09}.

The detection of the host galaxy allows us to subtract its contribution 
from the afterglow measured at $t \approx 36$\,d with ACS.  As can be seen from
Figure~\ref{fig:050820Ahst}, the host contribution at this epoch is 
significant and will affect the jet break time measured in 
\S~\ref{sec:models}. 

The combined X-ray, optical, and radio light curves of GRB\,050820A are shown
in Figure~\ref{fig:050820A}.

\subsection{GRB\,060418}
\label{sec:obs:060418}
GRB\,060418 was detected by the \Swift\ BAT at 03:06:08 \citep{GCN.4966}.
The $\gamma$-ray light curve shows three overlapping peaks with a total
duration $t_{90} = 52 \pm 1$\,s \citep{GCN.4975}.  GRB\,060418 was also bright 
enough to be detected with Konus-{\it Wind}, but no uncertainties were 
provided on the derived spectral parameters \citep{GCN.4989}.  We therefore 
use the rest-frame 1--10$^{4}$\,keV fluence derived from the \Swift-BAT by 
\citet{bkb+07}.  At $z=1.49$ \citep{pcb+07}, the total isotropic prompt energy 
release from GRB\,060418 was $E_{\gamma,\mathrm{iso}} = (1.0_{-0.2}^{+0.7}) 
\times 10^{53}$\,erg.

\begin{figure}
\centerline{\plotone{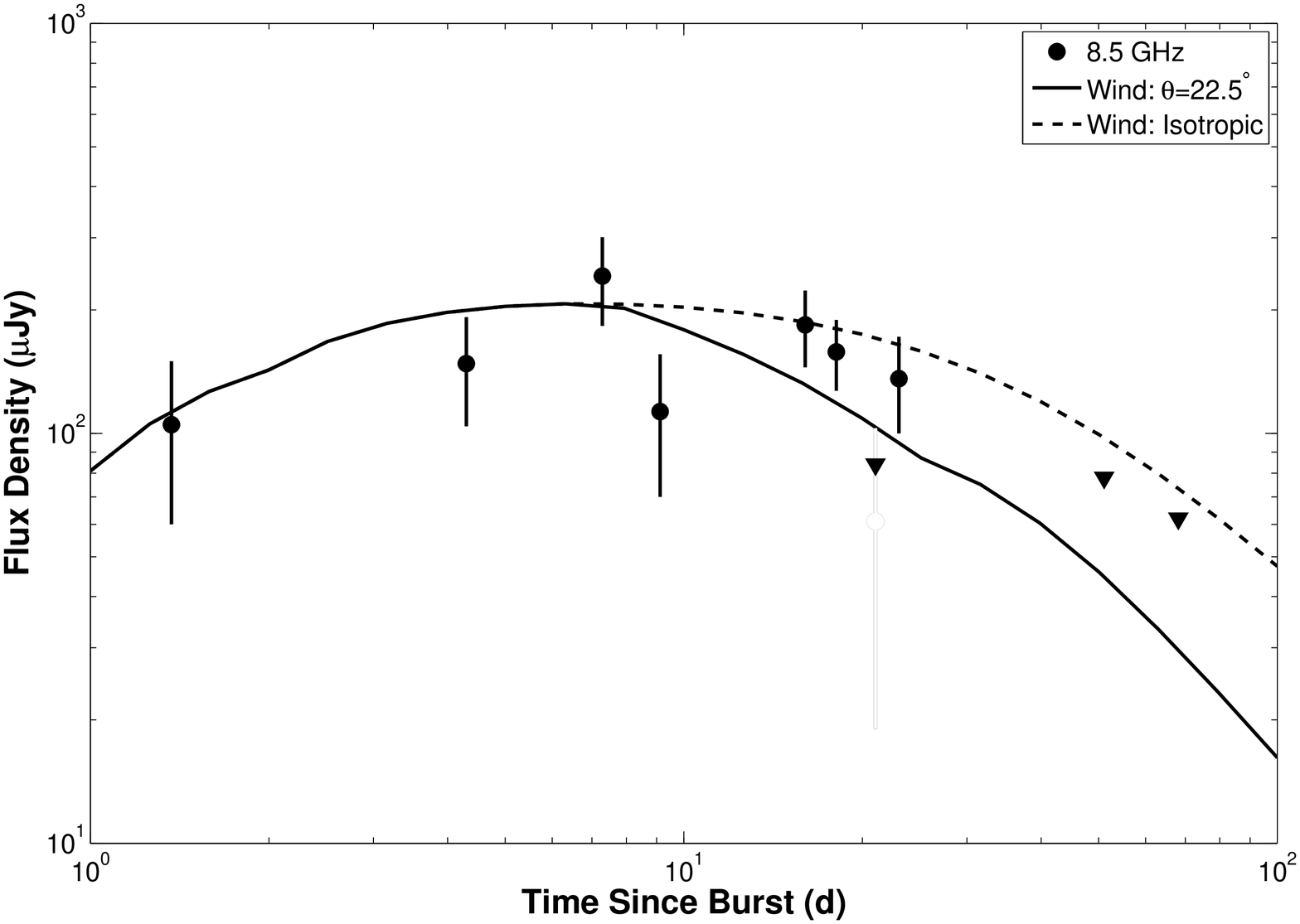}}
\caption[The X-ray, optical, and radio light curves of GRB\,060418]
  {The X-ray (previous page, top panel), optical (previous page, bottom panel), 
   and radio (this page) light curves of GRB\,060418.  The preferred
   model (Wind--Table~\ref{tab:060418}) is plotted as a solid line, while the 
   identical model for an isotropic explosion is shown as a dashed line.  The 
   optical data have been scaled by the model flux to match the \Rc-band.  The 
   optical shows a clear break at $t \approx 10$\,d, which is also favored in 
   the radio.  The X-ray afterglow is too faint to be detected at this time.}
\label{fig:060418}
\end{figure}

The XRT promptly slewed to the burst location and detected a fading X-ray
counterpart at $\alpha=15^{\mathrm{h}}45^{\mathrm{m}}42.8^{\mathrm{s}}$, 
$\delta=-03^{\circ}38\arcmin26\farcs1$ (J2000.0; $5\farcs8$ error radius;
\citealt{GCN.4966}).  Like many \Swift\ X-ray afterglows, the light curve
exhibits a bright flare at $t \approx 128$\,s super-posed on a power-law
decay \citep{GCN.4973}.  In Figure~\ref{fig:060418} we plot the X-ray
light curve evolution for $t > 0.1$\,d, obtained from the online
catalog of N.~Butler.

The automated Palomar 60\,inch (1.5\,m) telescope (P60; \citealt{cfm+06}) began
observing the afterglow of GRB\,060418 in the \Vc, \Rc, and \ip\ filters
beginning 2.7 hr after the burst (when the source became visible at Palomar
Observatory).  P60 data were reduced in the IRAF
environment using our custom real-time
reduction pipeline \citep{cfm+06}.  Where necessary, coaddition was performed
using Swarp\footnote{See http://terapix.iap.fr/soft/swarp.}.  Afterglow 
magnitudes were calculated with aperture photometry using an inclusion radius 
roughly matched to the full width at half-maximum intensity (FWHM) of the stellar point-spread function (PSF). Photometric calibration was performed 
relative to the calibration files provided by 
A.~Henden\footnote{Available via anonymous ftp at ftp.aavso.org.}, resulting 
in root-mean square (rms) variations of $\lesssim 0.05$\,mag in all filters. Photometric and 
instrumental errors have been added in quadrature to obtain the results 
presented in Table~\ref{tab:060418opt}.

Additional optical imaging was obtained with two large
ground-based facilities to supplement the P60 light curves at late times: 
the Large Format Camera (LFC) mounted on the 200\,inch (5.1\,m) Palomar
Hale telescope, and the Low Resolution Imaging Spectrometer (LRIS; 
\citealp{occ+95}) mounted on the 10\,m Keck I telescope.  All data were reduced 
in a manner similar to the P60 images using standard IRAF routines.   

\begin{figure}[t!]
  \centerline{\plotone{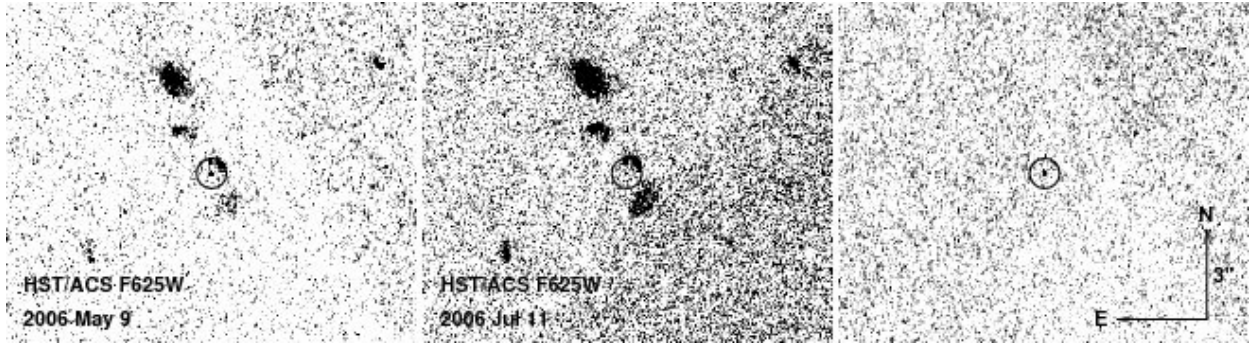}}
  \caption[\hst\ Imaging of GRB\,060418]
  {\hst\ Imaging of GRB\,060418.  \textit{Left:} An \textit{F625W} ACS image
   taken on 2006 May 9.  \textit{Center:} The same field in an image
   obtained on 2006 July 11.  \textit{Right:} A digital subtraction of the
   two images, revealing the residual afterglow emission.  There is no sign 
   of any host emission coincident with the afterglow; however the 
   photometry is complicated by contamination from several nearby sources 
   (which may be related to the host galaxy; \citealt{pcp+09}).}
\label{fig:060418hst}
\end{figure}

Late-time observations of GRB\,060418 were obtained with the Wide-Field Camera
(WFC) channel of the ACS on \hst\
(GO-10551; PI: Kulkarni).  The images were processed in an identical manner to 
\S~\ref{sec:obs:050820A}.  There is no evidence for host-galaxy emission 
directly coincident with the afterglow location.  However, several nearby sources, 
which may be related to the host galaxy \citep{pcp+09}, may 
contaminate the afterglow photometry.  The results of these observations are 
shown in Table~\ref{tab:060418opt} and Figure~\ref{fig:060418hst}.

The 1.3\,m Peters Automated Infrared Telescope (PAIRITEL; \citealt{bsb+06})
began observing the afterglow of GRB\,060418 at 5:25:34 on 2006 April 18.  
Full details of the PAIRITEL observations are presented by \citet{pcp+09}.  
Here we present the full multi-color PAIRITEL light curve of GRB\,060418, 
derived using aperture photometry and calibrated with respect to the
Two Micron All Sky Survey (2MASS; \citealt{scs+06}).  The results of our
analysis are shown in Table~\ref{tab:060418opt}.

Finally, we began observations of the fading optical counterpart of GRB\,060418
with the Very Large Array (VLA)\footnote{The VLA is operated by the National 
Radio Astronomy Observatory, a facility of the National Science Foundation 
operated under cooperative agreement by Associated Universities, Inc.} 
approximately 1 d after the burst.  The results of this and subsequent 
monitoring for 68 d after the burst are summarized in Table~\ref{tab:060418rad}.  
For the majority of the observations the antennae were in the A configuration;
the sole exceptions are the data points on 2006 June 8 (BnA) and 2006 June
25 (B).   All observations were reduced with the Astronomical Image Processing 
Software (AIPS) in the standard manner. 

\input{tab1}

The resulting X-ray, optical, and radio light curves of GRB\,060418 are shown
in Figure~\ref{fig:060418}.

\subsection{GRB\,080319B}
\label{sec:obs:080319B}
GRB\,080319B, or the ``naked-eye burst,'' has been discussed extensively in
the literature.  The most remarkable aspect of this unique event was the 
bright optical flash (peak optical magnitude of 5.3; \citealt{rks+08}) that
accompanied the prompt $\gamma$-ray emission.  Because of the temporal
coincidence, the contemporaneous $\gamma$-ray and optical emission are
believed to derive from the same physical region, with the observed 
$\gamma$-rays being generated by Compton scattering from the same
relativistic electrons that cause the optical flash \citep{rks+08,kp08}. 

The best constraints on the prompt emission come from the Konus-{\it Wind}
satellite.  The measured (20\,keV -- 7\,MeV) fluence does not change 
appreciably when converting to the standard bandpass.  The resulting 
isotropic prompt energy release at $z = 0.937$ \citep{rks+08,dfp+09} is 
$E_{\gamma,\mathrm{iso}} = (1.44 \pm 0.03) \times10^{54}$\,erg \citep{rks+08}.

For our modeling of GRB\,080319B, we draw on the rich data sets of 
\citet{rks+08}, \citet{bpl+09}, \citet{trl+08}, and \citet{ckh+09}.
To supplement these results, we obtained target-of-opportunity 
observations of the afterglow of GRB\,080319B in peak-up imaging mode
(i.e., with the blue 15.8\,$\mu$m filter) of the Infrared Spectrograph (IRS) 
on the \textit{Spitzer Space Telescope} through a Director's Discretionary 
Time proposal.  We obtained 60 dithered pointings, each consisting of 
two 30\,s cycles, beginning at 2008 March 21.81 ($\sim 2.55$\,d after the 
burst).  We clearly detect the afterglow at this time with a flux density of 
$35.7 \pm 3.9$\,$\mu$Jy \citep{GCN.7509}.  A second set of observations was 
obtained beginning at 2009 March 29.89, this time using 120 pointings.  
No source was detected to a (1$\sigma$) limiting magnitude of 4.0\,$\mu$Jy.  
The results of these observations are summarized in Table~\ref{tab:080319Brad}.

We also observed the location of GRB\,080319B with the Combined Array for 
Research in Millimeter-wave Astronomy (CARMA) at 95 GHz on 2008 March 20 
(mean time 11:30).  The configuration and data reduction are identical
to those described by \citet{ccf+08}.  We report a nondetection at the 
optical afterglow position with a 2$\sigma$ limit of 0.50\,mJy 
\citep{GCN.7493}.

GRB\,080319B was observed with the VLA at 4.8 and 8.5 GHz at two epochs in the 
first week after the burst \citep{GCN.7506}.  Both observations took place 
when the array was in the C configuration.  We also observed the afterglow
location for 33\,hr between 2008 December 20 and 2009 Jan 4 at 1.46\,GHz
to search for late-time emission.  The afterglow was not detected to a
2$\sigma$ upper limit of $< 28$\,$\mu$Jy.

\input{tab2}

The results of our millimeter and radio monitoring of GRB\,080319B are shown 
in Figure~\ref{fig:080319B} and Table~\ref{tab:080319Brad}, along with the 
previously published X-ray and optical light curves.

\section{Broadband Modeling Efforts}
\label{sec:models}
In the standard ``fireball'' formulation (e.g., \citealt{p05}), afterglow 
emission is powered by synchrotron radiation from relativistic electrons in 
the circumburst medium accelerated by an outgoing blast wave.  The resulting 
spectrum is well described as a series of broken power-laws with three 
characteristic frequencies: $\nu_{a}$, the frequency below which the radiation 
is self-absorbed; $\nu_{m}$, the characteristic frequency of the emitting 
electrons; and $\nu_{c}$, the frequency above which electrons are able to 
cool efficiently through radiation \citep{gs02}. 

The temporal evolution of the afterglow depends on the density profile
of the circumburst medium.  We consider here two possibilities: a constant
density circumburst medium [$\rho(r) \propto r^{0}$], as would be expected in 
an environment similar to the interstellar medium (ISM: \citealt{spn98}), and a wind-like environment [$\rho(r) 
\propto r^{-2}$], as would be the case for a massive star progenitor shedding 
its outer envelope at a constant rate before core collapse
\citep{cl00}.

GRBs are believed to be highly collimated explosions \citep{r99,sph99}.
At early times, observers only notice emission from a narrow cone (opening 
angle $\theta \approx \Gamma^{-1}$, where $\Gamma$ is the Lorentz factor of the 
expanding shock) due to relativistic beaming.  The resulting evolution therefore
mimics an isotropic explosion.  As the shock slows,
however, lateral spreading of the jet becomes important, and the observer
eventually notices ``missing'' emission from wider angles.
This hydrodynamic transition manifests itself as an achromatic steepening
in the afterglow light curve.  Measuring the time of this jet break 
($t_{\mathrm{j}}$) allows us to infer the opening angle of the outflow 
($\theta$).

Our objective here is to translate the observed three critical frequencies, 
together with the peak flux density, $F_{\nu,\mathrm{max}}$, and the jet break 
time, $t_{\mathrm{j}}$, into a physical description of the outflow.  In
particular, we shall attempt to estimate seven parameters: $E_{\mathrm{KE}}$, the
kinetic energy of the blast wave; $n$, the density of the circumburst medium;
$\epsilon_{e}$, the fraction of the total energy apportioned to electrons;
$\epsilon_{B}$, the fraction of the total energy apportioned to the magnetic
field; $p$, the electron power-law index; $A_{V}$, the host-galaxy extinction;
and $\theta$, the jet opening angle.  We make use of the software
described by \citet{yhs+03}, a multi-parameter fitting program incorporating
the standard afterglow formulation, as well as corrections for radiative
losses and inverse-Compton emission \citep{se01}.

To account for differences in instrumental configurations, we have applied a 
5\% cross-calibration uncertainty to all data points before calculating the 
models.  All reported uncertainties have been determined using a Markov-Chain 
Monte-Carlo analysis with 1000 trials and represent only statistical errors 
associated with the fit.  Systematic errors associated
with model uncertainties are likely much larger and difficult to estimate.

\subsection{GRB\,050820A}
\label{sec:models:050820A}
The optical and X-ray light curves of GRB\,050820A exhibit a dramatic 
rebrightening at $t \approx 220$\,s, both jumping in concert with a strong 
rise in the $\gamma$-ray emission \citep{vww+06,ckh+06}.  We therefore remove 
all X-ray and optical points at early times ($t < 0.1$\,d) from our fitting 
routines.  Likewise, the radio light curve exhibits a bright flare at $t 
\approx 1$\,d that is probably due to reverse shock emission \citep{ckh+06}.  
Since our modeling software only includes contributions from the forward 
shock, we include only radio observations at $t > 5$\,d in our models.

\begin{figure}
\centerline{\plotone{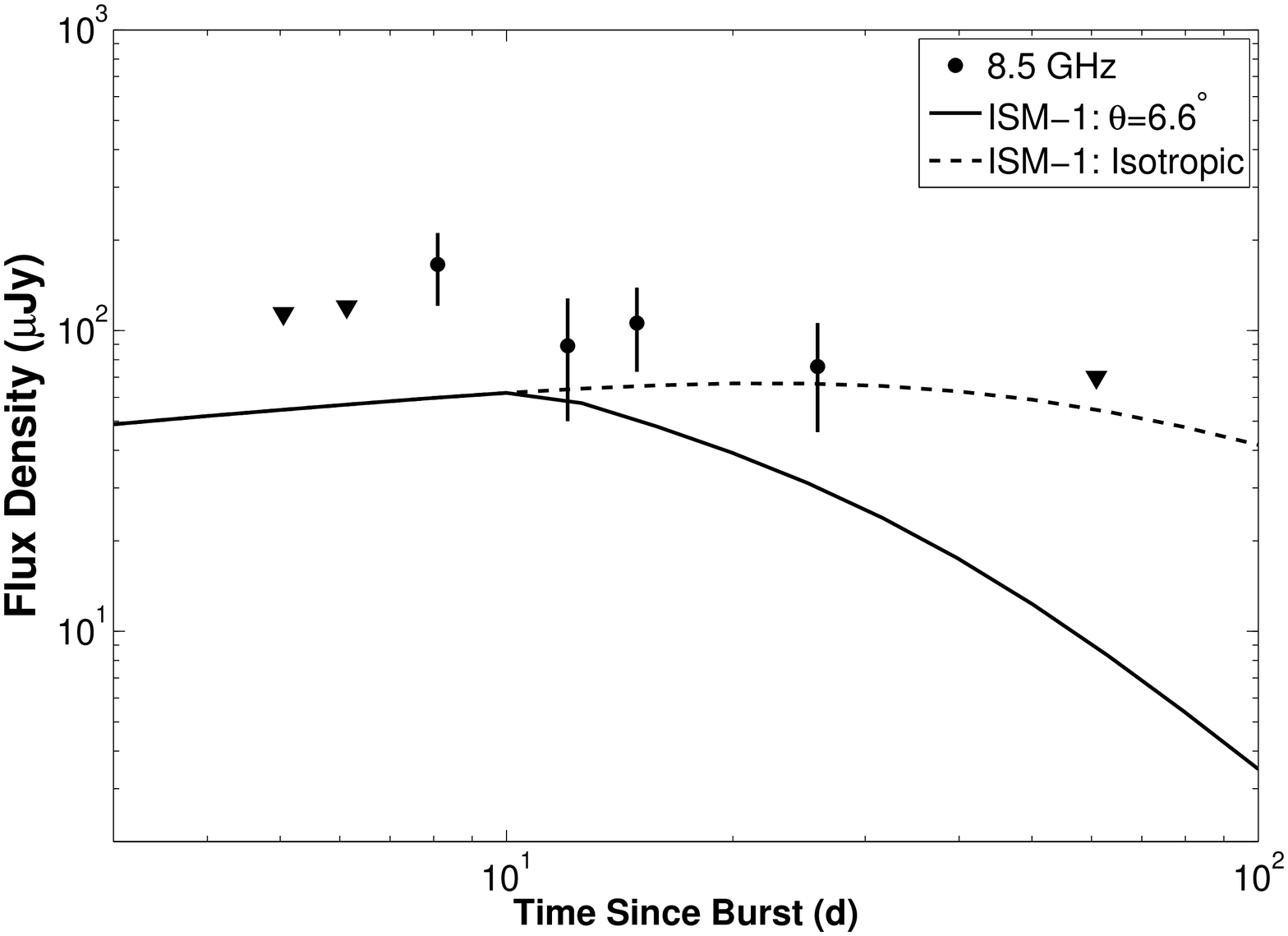}}
\caption[The X-ray, optical, and radio light curves of GRB\,050820A]
  {The X-ray (previous page, top panel), optical (previous page, bottom panel), 
   and radio (this page) light curves of GRB\,050820A.  Radio observations
   at $t < 5$\,d are left out of the model, as the emission at this time is
   likely dominated by the reverse shock \citep{ckh+06}.  The preferred
   model (ISM-1; Table~\ref{tab:050820A}) is plotted as a solid line, while 
   the identical model for an isotropic explosion is shown as a dashed line.  
   The optical data have been scaled by the model flux to match the \Rc-band.  
   Both the X-ray and optical bandpasses show a clear break at $t \approx 10$\,d.  
   The radio is not sufficient to distinguish between an isotropic and a 
   collimated outflow.}
\label{fig:050820A}
\end{figure}

The best-fit afterglow models for GRB\,050820A are plotted in 
Figure~\ref{fig:050820A}, and the relevant physical parameters are provided in 
Table~\ref{tab:050820A}.  As discussed by \citet{ckh+06}, the optical light 
curve exhibits a distinct break between the last ground-based optical detection 
at $t \approx 7$\,d and the \hst\ observations at $t \approx 36$\,d.  The 
additional late-time X-ray data firmly establish the presence of this break 
in the X-ray bandpass as well, cementing the explanation as a jet break.  The 
radio data at late times are not sufficient to distinguish between a beamed 
and isotropic outflow.

\input{tab3}

For a constant density circumburst medium, we find two solutions with somewhat
different model parameters but similar overall fit quality, and both are 
presented in Table~\ref{tab:050820A}.  Though values of the electron index 
$p$ less than 2 require a somewhat artificial cutoff to keep the total 
energy carried by the electrons finite, we slightly prefer the model with 
$p = 1.75$ (see, e.g., \citealt{b01} for a discussion of afterglows with 
electron index $p < 2$).  Here the density scale is more in line with our 
expectation that long-duration GRBs should inhabit dense regions of recent 
star formation, as well as previous GRB afterglow modeling results (e.g., 
\citealt{pk02,yhs+03}).  Though the fraction of the total energy partitioned 
to the magnetic field, $\epsilon_{B}$, is small, comparable values have been 
inferred for several other previous GRBs (e.g., \citealt{pk02}).

Most importantly for our purposes, the kinetic energy of the afterglow, 
$E_{\mathrm{KE}}$,  and the jet opening angle, $\theta$, are relatively similar 
in the two models.  Though both require extremely large isotropic kinetic 
energies, the values inferred here are only an order of magnitude larger than 
the prompt $\gamma$-ray energy release.  This modest $\gamma$-ray conversion 
efficiency ($\eta_{\gamma} \equiv E_{\gamma} / [E_{\gamma} + E_{\mathrm{KE}}] \sim 
10$\%) is consistent with theoretical predictions of the internal shock 
model \citep{kps97,dm98}.

The jet break time occurs somewhat earlier than originally suggested by 
\citet{ckh+06}.  This is due to a combination of improved constraints from 
the X-ray afterglow coupled with a shallower post-break decay index inferred 
for the $p=1.75$ model.

We were unable to obtain any high-quality fits to the afterglow of GRB\,050820A
assuming a wind-like circumburst medium.  We shall return to the lack of 
evidence for massive-star progenitors from GRB afterglow modeling in 
\S~\ref{sec:conclusion}.

\subsection{GRB\,060418}
\label{sec:models:060418}
The most striking feature in the light curve of GRB\,060418 is a bright X-ray 
flare at $t \approx 300$\,s \citep{GCN.4973}.  Such flares have been reported 
in a large fraction of \Swift\ XRT light curves \citep{fmr+07}, and are widely 
believed to be caused by late-time energy injection from the central engine 
\citep{zfd+06}.  These X-ray flares can in some cases contribute a significant 
fraction of the prompt energy release to the total energy budget, and 
therefore have a large effect on the post-flare decay \citep{fmr+07}.  Rapid 
variability in the X-ray light curve of GRB\,060418, inconsistent with 
standard afterglow models, is seen as late as several hours after the burst.  
Like GRB\,050820A, we therefore only include observations at $t > 0.1$\,d in 
our broadband modeling analysis.

The resulting fits and best-fit parameters are shown in 
Figure~\ref{fig:060418} and Table~\ref{tab:060418}.  Again a clear break is 
seen in the optical light curve at late times ($t \approx 8$\,d).  The X-ray 
afterglow has dropped below the XRT threshold at this time.  However, the 
radio afterglow is still detected and exhibits some evidence for a steepening 
decay consistent with that seen in the optical.  Unfortunately, the break
time is not very well constrained, either in the optical or radio bands. 
Though not entirely conclusive, we consider this relatively strong evidence 
in support of collimation.      

\input{tab4}

Unlike most previously modeled afterglows (c.f., GRB\,050904; 
\citealp{fck+06}), our results indicate that the electron cooling frequency, 
$\nu_{c}$, fell below the optical bands over the duration of our observations.  
The forward-shock emission above $\nu_{c}$ is independent of the circumburst 
medium profile, leading to indistinguishable fits in the X-ray and optical 
bandpasses.  While the radio behavior is divergent at early  and late times, 
our observations are not sufficient to conclusively distinguish between the 
two models.  Since the wind model provides a slightly better fit and does not 
require microphysical parameters held fixed at or near equipartition, we 
shall adopt it for the remainder of the work.

The primary drawback of the wind-like scenario, however, is the extremely high 
$\gamma$-ray efficiency.  Somehow the physical process generating
the prompt emission must have been capable of converting $\sim 99\%$ of
the outgoing blast-wave energy to $\gamma$-rays, while most internal shock
models predict a maximum $\gamma$-ray efficiency of $\sim 10\%$ 
\citep{kps97,dm98}.  We shall return to this issue in \S~\ref{sec:energy}.

\subsection{GRB\,080319B}
\label{sec:models:080319B}
Several groups (e.g., \citealt{rks+08,bpl+09,wvp+09}) have presented detailed
observations of the early afterglow of GRB\,080319B, revealing a complex 
behavior not easily understood in the context of the standard fireball model.  
In particular we note that the optical spectral index, $\beta_{\rm O}$, evolves 
dramatically at early times, and as late as 0.5\,d after the burst is too 
shallow to be accommodated by our forward shock models ($\beta_{O} \lesssim 
0.2$; \citealt{bpl+09}).  We therefore consider the evolution of the 
afterglow only at $t > 0.5$\,d.  

In addition, the late-time ($t \gtrsim 10$\,d) optical light curve exhibits a 
pronounced red bump not seen in the X-rays that has been attributed to 
emission from an underlying supernova \citep{trl+08,bpl+09}.  Such features 
have now been seen in many relatively nearby GRB optical afterglows 
\citep{zkh04}, and are not accounted for in our synchrotron formulation.  We 
therefore leave all $\rp$, $\ip$, and $\zp$ measurements out of our models 
at $t > 10$\,d.

\begin{figure}
\centerline{\plotone{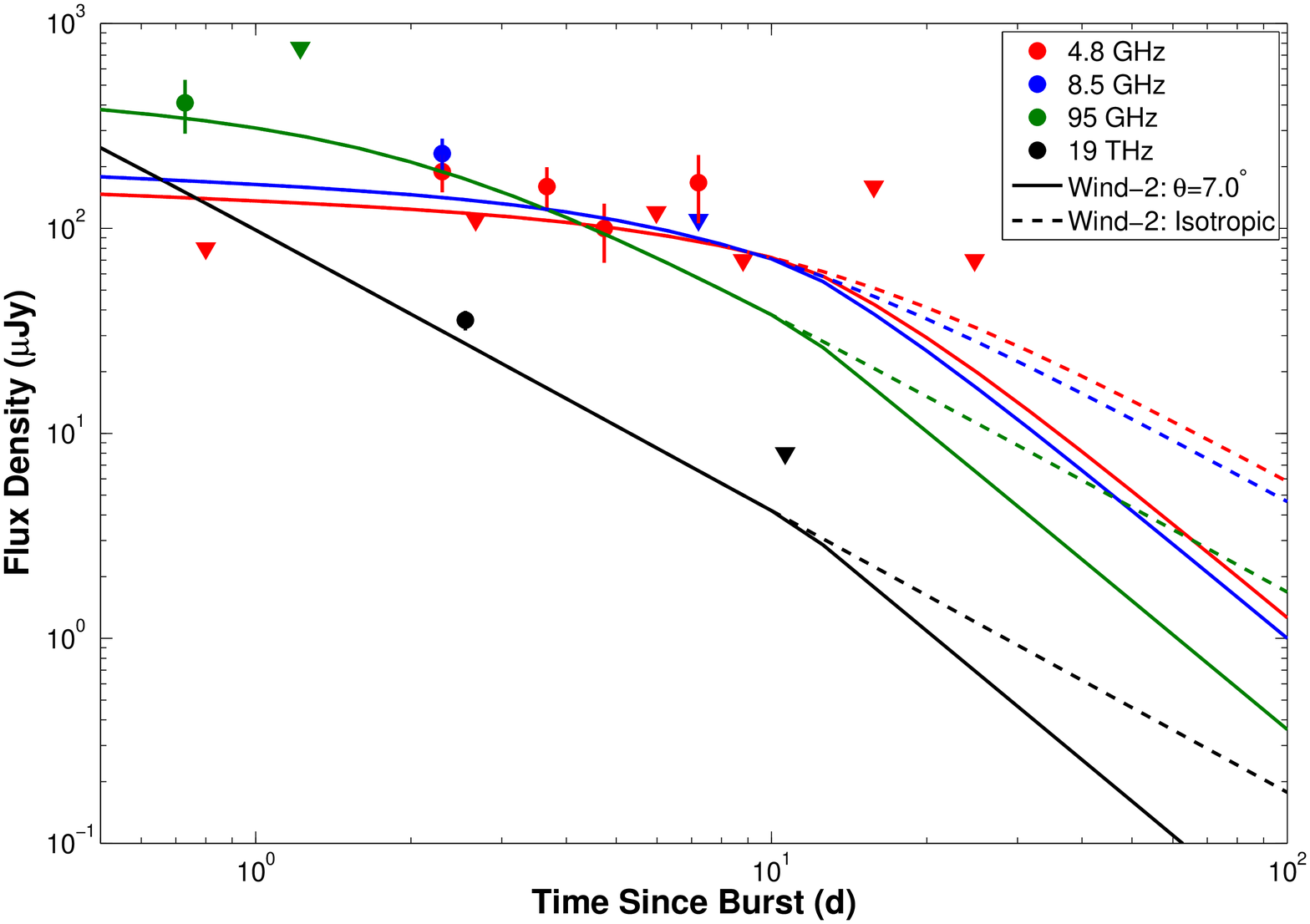}}
\caption[The X-ray, optical, and radio light curves of GRB\,080319B]
  {The X-ray (previous page, top panel), optical (previous page, bottom panel), 
   and radio (this page) light curves of GRB\,080319B.  The preferred
   model (Wind-2--Table~\ref{tab:060418}) is plotted as a solid line, while the 
   identical model for an isotropic explosion is shown as a dashed line.  Both 
   the X-ray and optical show a break at $t \approx 10$\,d.  The radio afterglow 
   is too faint at this time to see evidence for collimation.  We have left 
   out the late-time ($t > 10$\,d) \rp, \ip, and \zp\ data from our fits due 
   to the presence of contaminating SN emission.}
\label{fig:080319B}
\end{figure}

The resulting fits are plotted in Figure~\ref{fig:080319B}, with the model 
parameters presented in Table~\ref{tab:080319B}.  As was pointed out by 
\citet{rks+08} and most convincingly by \citet{trl+08}, both the X-ray and 
optical light curves exhibit a break at $t \approx 12$\,d.  The break is 
only clearly visible in the redder optical bands after including the 
contribution from the underlying supernova.  Here the radio observations are 
unable to provide any constraints on the presence of a jet break, as the 
radio afterglow was comparably faint and only detected over the first week 
of observations.

\input{tab5}

We find that two models provide a reasonable fit to the broadband data, and both 
require a wind-like circumburst environment.  The parameters derived for 
the first model, with $p = 2.10$, are broadly similar to those derived for 
the ``wide'' jet (see below) by \citet{rks+08}, with one notable exception: 
$\epsilon_{B}$ differs by two orders of magnitude (fixed at equipartition in 
our model, compared with $3 \times 10^{-3}$ in \citealt{rks+08}).  Taken 
together, we find that the model parameters provided in \citet{rks+08} provide a 
relatively poor fit to the late-time data, particularly in the optical bands.

An alternate model, with $p = 1.85$, provides a marginally better fit to 
the data.  Since the inferred density is slightly larger and more in line 
with previous GRB afterglows, we adopt this as our preferred model for the 
afterglow of GRB\,080319B.  We note again that the discrepancy between the 
two competing models is relatively small with respect to the afterglow 
energy and opening angle.

\citet{rks+08} have incorporated the early-time data into their model of 
GRB\,080319B by invoking a double-jetted system: the high-energy emission is 
focused in a narrow ($\theta \approx 0.2^{\circ}$) jet that dominates the early 
afterglow ($t \lesssim 0.5$\,d), while the material at lower Lorentz factor 
powering the late-time afterglow is channeled into a wider jet ($\theta \approx 
4^{\circ}$).  Such a configuration is not without precedent and has been 
invoked to explain multiple breaks in the light curve of GRB\,030329 
\citep{bkp+03}.  Explaining the early emission from GRB\,080319B is beyond 
the scope of this work; however, we consider the implications of multiple-jet 
models in \S~\ref{sec:conclusion}.

\subsection{GRBs\,050904 and 070125}
\label{sec:models:other}
Finally, for completeness we include here a brief summary of the primary 
results from the modeling of GRB\,050904 \citep{fck+06} and GRB\,070125 
\citep{ccf+08}.  

To date, GRB\,050904 is the third most distant spectroscopically confirmed GRB ($z=6.295$;
\citealt{kka+06,hnr+06}).  The optical and X-ray light curves exhibit a 
prominent break at $t = 2.6 \pm 1.0$\,d \citep{tac+05}, resulting in an 
opening angle of $\theta \approx 8^{\circ}$.  The afterglow was notable for 
an extremely large inferred density: $n \approx 700$\,cm$^{-3}$ for an 
isotropic circumburst medium.  Even after applying a collimation correction, 
the total energy release from GRB\,050904 was in excess of  $10^{52}$\,erg, 
making it one of the most energetic explosions ever detected.

Although \Swift\ did not trigger immediately on GRB\,070125, the $\gamma$-ray
emission was bright enough to be detected by both Konus-{\it Wind} and 
{\it RHESSI}, providing superb coverage of the high-energy properties of this 
event \citep{bhp+08}.  The radio afterglow of GRB\,070125 was one of the 
brightest in the \Swift\ era, making it an ideal source for broadband 
modeling.  A clear break is seen in the optical light curve at $t 
\approx 4$\,d \citep{GCN.6096}.  While the X-ray light curve also undergoes
a steepening around this time, it occurs slightly later than in the optical 
bandpass.  This may be due to the effects of inverse-Compton emission 
dominating the X-ray afterglow at this time.  The circumburst density inferred 
for GRB\,070125 was also relatively high ($n \approx 40$\,cm$^{-3}$), 
resulting in a strongly self-absorbed radio spectrum.  

The relevant energy properties of all five of the events in our sample are 
summarized in Table~\ref{tab:energy}.

\input{tab6}

\section{Discussion}
\label{sec:energy}
In the previous section, we provided model fits to the broadband afterglows 
of five \Swift\ GRBs, all of which exhibit evidence for a collimated, 
relativistic outflow.  In some cases, the breaks are clearly visible across 
the X-ray, optical, and radio bandpasses (e.g., GRB\,070125), while in others 
the data are insufficient to verify the achromatic nature of the observed 
break (e.g., the X-rays for GRB\,060418).  Regardless, the fact that all five 
events are consistent with a beamed outflow is in marked contrast to previous 
searches for jet break candidates in \Swift\ events 
\citep{p07,kb08,rlb+08,lrz+08}.

The most natural explanation for this discrepancy is the role of selection 
effects.  As is evident from Figure~\ref{fig:eiso}, the heightened 
sensitivity of \Swift\ is preferentially selecting GRBs with smaller 
$E_{\gamma,\mathrm{iso}}$ values compared with pre-\Swift\ missions.  Bandpass 
likely exacerbates this effect: the observed correlation between the peak 
energy of the prompt $\gamma$-ray spectrum ($E_{p}$) and 
$E_{\gamma,\mathrm{iso}}$ \citep{a06} further suggests that \Swift\ is detecting an 
underluminous sample with respect to previous missions with extended 
high-energy coverage.

The result is that \Swift\ jet breaks should occur on average later than 
pre-\Swift\ events, making them more difficult to observe (for a given 
sensitivity limit).   Such an effect was predicted (in the context of the 
structured jet model) by \citet{psf03}. GRB\,060418 offers an illustrative 
example; with $E_{\gamma,\mathrm{iso}} = 10^{53}$\,erg, it falls in 
the 80th percentile of the \Swift\ $E_{\gamma,\mathrm{iso}}$ distribution, and 
with $t_{j} = 7.6$\,d, the break occurred at $R \sim 24.5$\,mag.  The X-rays 
were already too faint at this time to be detected, and the only reason for 
the optical detection was the deep \hst\ imaging.  Given the typical follow-up 
capabilities of a medium-aperture telescope, jet breaks are virtually 
undetectable for a majority of \Swift\ GRBs \citep{dgp+08,kb08}.  In fact,
some of the faintest events (in terms of $E_{\gamma,\mathrm{iso}}$) may be 
isotropic and still consistent with our observed energetics distribution.  

\begin{figure}[t!]
  \centerline{\plotone{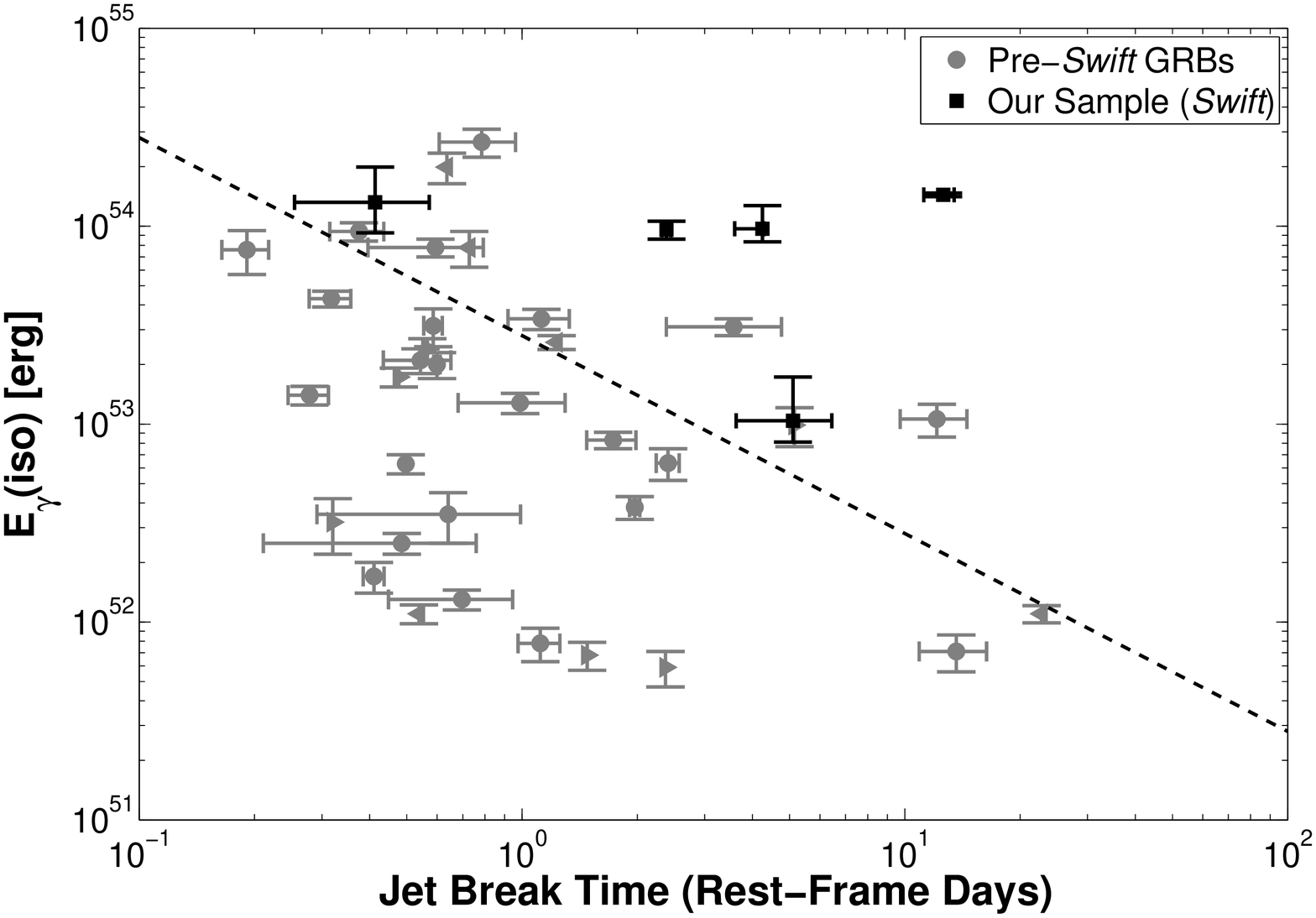}}
  \caption[Afterglow jet break time (rest frame) as a function of 
   $E_{\gamma,\mathrm{iso}}$.]
  {Afterglow jet break time (rest frame) as a function of 
   $E_{\gamma,\mathrm{iso}}$.  The dashed line represents a roughly constant 
   collimation-corrected prompt energy release (there is weak dependence on
   the circumburst density and $\gamma$-ray efficiency).  The events in our 
   sample have on average larger jet break times than analogous pre-\Swift\ 
   events.  This result suggests that the $E_{\gamma}$ distribution is wider than
   previously thought in a relatively model-independent manner.  Pre-\Swift\
   events have been compiled from \citet{fb05}.}
\label{fig:tjet}
\end{figure}

In Figure~\ref{fig:tjet}, we plot the observed jet break times for our sample 
as a function of $E_{\gamma,\mathrm{iso}}$ compared to the pre-\Swift\ sample 
from \citet{fb05}.  The solid line indicates a constant collimation-corrected
prompt energy release of $E_{\gamma,\mathrm{iso}} \approx 10^{51}$\,erg.  We wish
to emphasize that the derived collimation angle is only weakly dependent
on two model parameters: $n$ and $\eta_{\gamma}$ (both to the $1/8$ power).
As a result, though the afterglow energy may be relatively uncertain, it is 
clear that the $E_{\gamma}$ values derived for the events in our sample will be 
significantly larger than the typical pre-\Swift\ GRB.

\begin{figure}[t!]
  \centerline{\plotone{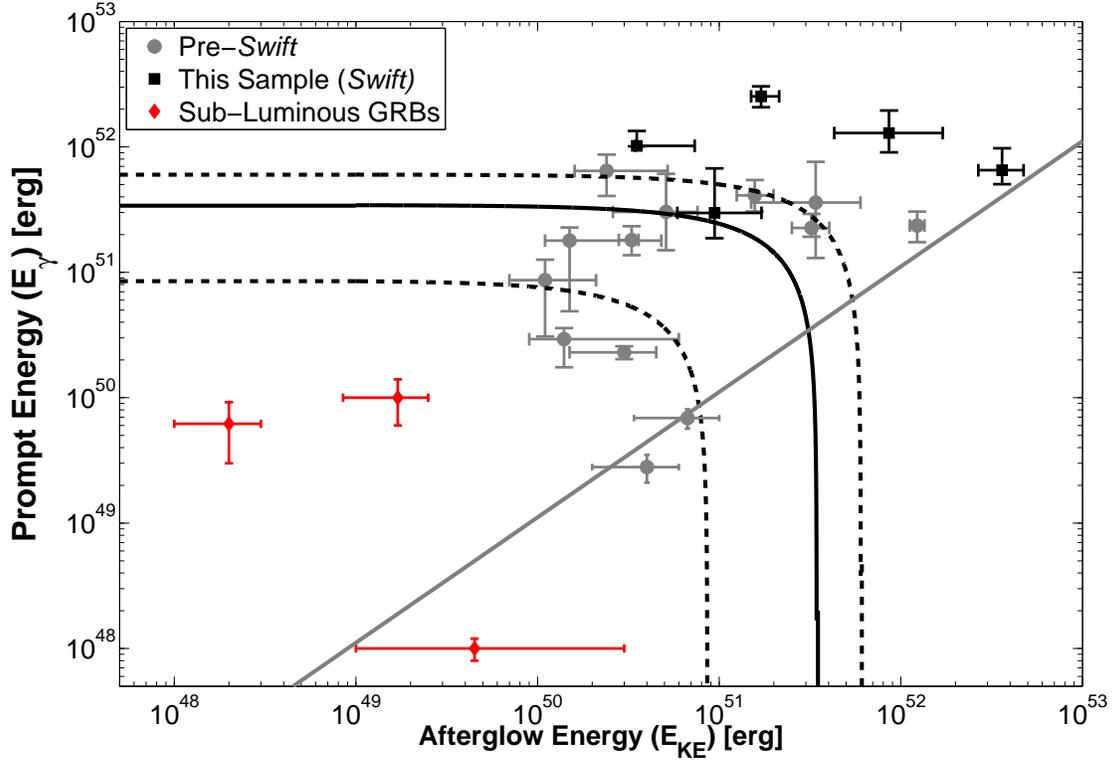}}
  \caption[The two dimensional ($E_{\mathrm{KE}} + E_{\gamma,\mathrm{iso}}$) 
   relativistic energy release from long-duration GRBs.]
  {The two dimensional ($E_{\mathrm{KE}} + E_{\gamma,\mathrm{iso}}$) relativistic 
  energy release from long-duration GRBs.  Neglecting the most nearby,
  underluminous GRBs (GRBs\,980425, 031203, and 060218), pre-\Swift\ events 
  are clustered around a total energy release of $\approx 3 \times 
  10^{51}$\,erg (solid line; dashed lines show the rms of the distribution).  
  With the exception of GRB\,060418, all of the events in our sample fall well 
  outside this distribution, typically with $E_{\mathrm{total}} \gtrsim 
  10^{52}$\,erg.  The solid grey line reflects a constant $\gamma$-ray
  efficiency of $\eta_{\gamma} = 0.1$.  That so many of the observed events 
  fall above this line suggests internal shocks may not be responsible for 
  the generation of the prompt $\gamma$-ray emission.  References -- 
  \citet{pk02}: GRBs\,990123, 990510, 991208, 991216, 000301C, 010222;
  \citet{yhs+03}: GRBs\,970508, 980703, 000926; \citet{bkf04}: GRBs\,970508,
  980703; \citet{clf04}: GRB\,020405; \citet{bdf+01}: GRB\,000418;
  \citet{bkp+03}: GRB\,030329; \citet{skb+04}: GRB\,020903; \citet{lc99}:
  GRB\,980425; \citet{skb+04}: GRB\,031203; \citet{skn+06}: GRB\,060218.}
\label{fig:energy}
\end{figure}

The final, collimation-corrected energy release from the events in our sample, 
both from the prompt $\gamma$-ray emission and powering the afterglow, is 
plotted in Figure~\ref{fig:energy}.  Also plotted are analogous results from 
previous studies of pre-\Swift\ afterglows (see figure caption for references). 
With the exception of the most nearby events (red diamonds), the total energy 
release ($E_{\gamma} + E_{\mathrm{KE}}$) from pre-\Swift\ GRBs was relatively 
tightly clustered around a value of $\sim 3 \times 10^{51}$\,erg (solid line in 
Figure~\ref{fig:energy}).  Clearly the GRBs presented here are inconsistent 
with this distribution, preferentially falling at the high-energy end.  
Several events exceed $10^{52}$\,erg in total energy release, something only 
achieved for a single event in the pre-\Swift\ era (GRB\,970508; 
\citealt{yhs+03,bkf04}).  Much as has been seen at the low-energy end,
our results suggest that the true energy release from GRBs is relatively broad 
and capable of extending out to at least $10^{52}$\,erg.

Finally, we return to the issue of $\gamma$-ray efficiency.  As has also been seen 
in the pre-\Swift\ era, the inferred $\gamma$-ray efficiency can often 
be dramatically higher than would be predicted from the internal shock model 
($\eta_{\gamma} \lesssim 0.10$; \citealt{kps97,dm98}).  The problem may be 
eased to some extent by the nature of our model parameters, as 
$E_{\mathrm{KE}}$ is measured at the time of the transition from fast to slow 
cooling, after which the shock may have lost a significant fraction of its 
initial energy \citep{ccf+08}.  However, this would necessarily increase the 
total energy budget, possibly at times to values approaching $10^{53}$\,erg.  
An alternative possibility is that the $\gamma$-rays are produced via 
relativistic turbulence \citep{lnp09,nk09}, where the $\gamma$-ray efficiency
can approach unity.  Such a model has been already been invoked for 
GRB\,080319B \citep{kn09}, and clearly merits further study.

Alternatively, if the flux distribution were dominated by small-scale 
fluctuations (of angular scale $\Gamma^{-1}$), this may cause a large
variation in the energy distribution (the ``patchy shell'' model; 
\citealt{kp00}).  This would have a particularly strong influence on 
$E_{\gamma}$, as the energy would be averaged over a much smaller physical
scale, and could lead to large fluctuations in the value of $\eta_{\gamma}$.  

\section{Conclusions}
\label{sec:conclusion}
In this work we have presented model fits to the broadband afterglows of the 
\Swift\ GRBs\,050820A, 060418, and 080319B.  Together with previous results 
from GRB\,050904 and GRB\,070125, we demonstrate all five events are consistent 
with our understanding of relativistic, collimated explosions.  However, the 
inferred opening angles for several events are larger than would be expected 
if GRBs were truly standard candles.  The result is a broad 
collimation-corrected energy distribution, with some events emitting in 
excess of $10^{52}$\,erg.

It is at first glance somewhat surprising that \Swift\ has discovered a large 
fraction of the most energetic GRBs.  This may be due in large part to 
sample size.  As can be seen from Figure~\ref{fig:eiso}, nearly all the GRBs 
in our sample (except GRB\,060418) were brighter (in terms of 
$E_{\gamma,\mathrm{iso}}$) than nearly all of the pre-\Swift\ GRBs having known 
redshifts.  Even though the median $E_{\gamma,\mathrm{iso}}$ value of \Swift\ 
bursts is smaller than the pre-\Swift\ sample, the increase in the total 
number of GRBs detected by \Swift\ has enabled it to target the extreme ends of 
the energy distribution.  These events must be relatively rare,
or they would have been easily detected by previous high-energy missions.

Despite their rarity, the most energetic events provide some of the strongest 
constraints on possible progenitor models.  The maximum energy release in
magnetar models \citep{u92} is $\sim 3 \times 10^{52}$\,erg, which is set
by the rotational energy of a maximally rotating neutron star (see, e.g.,
\citealt{tcq04,mtq07}).  The fact that several events in our sample approach
this limit suggests that the central engine remnant should be a black hole 
(at least for these hyper-energetic GRBs).  A lingering question, then, is 
how to produce the long-lived ($t \gg \Delta t_{\mathrm{GRB}}$) central engine 
activity seen commonly in \Swift\ X-ray afterglows.  While this could 
naturally be explained in the magnetar model, the mechanism for generating 
this activity from black holes is less clear.  Though various theories have 
been proffered (e.g., \citealt{zfd+06}), this remains a vexing issue.

Just as troubling is the lack of evidence for progenitor mass loss in the 
density profile of the circumburst medium of some events.  In the case of 
radio supernovae, where the shock expansion is Newtonian, the circumburst 
medium is consistently well fit with an $r^{-2}$ density profile (e.g., 
\citealt{sck+06}).  That this is not the case for long-duration GRBs, given 
the preponderance of evidence for their association with massive stars 
(\citealt{wb06} and references therein), should give pause in over-interpreting the parameters derived 
from these models.

One possible explanation may be our relatively poor understanding of the 
final stages of stellar evolution.  A variety of recent results, ranging from 
the discovery of fast-moving ($v \approx 6000$\,km s$^{-1}$) material ejected 
from $\eta$ Car \citep{s08}, to the dense circumstellar material partially powering the
luminous ($M_{V} \approx -22$\,mag) SN\,2006gy \citep{ock+07,slf+07}, to
detection of a pre-SN outburst from SN\,2006jc \citep{psm+07,fsg+07} and
SN\,2005gl \citep{gl09}, suggest that massive stars undergo violent periods 
of episodic mass loss in the late stages of stellar evolution.  Future 
theoretical progress to pin down the expected density profile surrounding 
massive stars in the latest stages of stellar evolution at distances relevant 
to GRB afterglows ($\sim 1$\,pc) might help to resolve this discrepancy 
(e.g., \citealt{rgs+05,d07}).

An additional concern, motivated by the double-jet models for GRB\,030329
\citep{bkp+03} and GRB\,080319B \citep{rks+08},
is our assumption that the entire relativistic outflow is collimated into a 
uniform jet with a single opening angle (the so-called ``top-hat'' model).  
A variety of other models for structured jets have been proposed, typically 
with the Lorentz factor of the outflow varying as a function of angle from 
the jet axis (see, e.g., \citealt{g07} for a review).  Though the double-jet 
model is no doubt a bit contrived, theoretical simulations of relativistic 
jets suggest the top-hat model is overly simplistic (e.g., 
\citealt{zwh04,tmn08}).  It would 
also ease the sometimes extreme efficiency requirements if the $\gamma$-ray 
emission were more narrowly beamed than the afterglow.  However, such models 
imply the existence of so-called ``on-axis'' orphan afterglows, where the
line of sight misses the $\gamma$-ray emission but an observer still sees a
regular (on-axis) afterglow.  The lack of such events to date limits the 
X-ray to $\gamma$-ray beaming factor ratio to $< 2:1$, while future 
wide-field, high-cadence surveys will soon do the same for the optical 
\citep{np03}.

Finally, we consider the future prospects for the study of the most energetic 
GRBs.  As we suggested previously \citep{ccf+08}, the recent launch of the 
{\it Fermi} satellite offers an incredible opportunity in this respect.  The 
high-energy bandpass of the Large Area Telescope, extending out to hundreds 
of GeV, is ideally suited to target the bright end of the 
$E_{\gamma,\mathrm{iso}}$ distribution.  Already in less than a year of 
operation, {\it Fermi} has detected two of the brightest GRBs ever, 
GRB\,080916C \citep{aaa+09,gck+09} and GRB\,090323 \citep{GCN.9021,GCN.9030}, 
with $E_{\gamma,\mathrm{iso}}$ in excess of $5 \times 10^{54}$\,erg.  Coupled 
with the large Lorentz factor required to produce GeV photons, 
multi-wavelength campaigns targeted at such events are well positioned to 
search for early jet breaks when the afterglow is still bright.  Together, 
synergistic \Swift\ and {\it Fermi} observations in the coming years should 
be able to shed a good deal of light on the high end of the GRB energy
distribution.

\acknowledgments{
S.B.C.~and A.V.F.~wish to acknowledge generous support from Gary and Cynthia 
Bengier, the Richard and Rhoda Goldman fund, and National Science Foundation (NSF)
grant AST--0607485.  A.G.~acknowledges support by the Israeli Science Foundation,
an EU Seventh Framework Programme Marie Curie IRG fellowship and the 
Benoziyo Center for Astrophysics, a research grant from the Peter and 
Patricia Gruber Awards, and the William Z.~and Eda Bess Novick New
Scientists Fund at the Weizmann Institute.  J.N.N.~is supported by
National Aeronautics and Space Administration (NASA)
contract NAS5-00136.  T.P.~acknowledges support from an ERC advanced research 
grant.  P60 operations are funded in part by NASA through the 
{\it Swift} Guest Investigator Program (grant number NNG06GH61G).
Based on observations made with the NASA/ESA {\it Hubble Space Telescope}, obtained 
from the Data Archive at the Space Telescope Science Institute, which is 
operated by the Association of Universities for Research in Astronomy, Inc., 
under NASA contract NAS 5-26555. These data are associated with 
program GO-10551.  This work is based in part on observations made with the 
{\it Spitzer Space Telescope}, which is operated by the Jet Propulsion Laboratory, 
California Institute of Technology under a contract with NASA.
This publication has made use of data obtained from the \Swift\ interface
of the High-Energy Astrophysics Archive (HEASARC), provided by
NASA's Goddard Space Flight Center.  Support for CARMA construction was 
derived from the Gordon and Betty Moore Foundation, the Kenneth T. and 
Eileen L. Norris Foundation, the Associates of the California Institute of 
Technology, the states of California, Illinois, and Maryland, and the NSF. 
Ongoing CARMA development and operations are supported by 
the NSF under a cooperative agreement, and by the 
CARMA partner universities.  PAIRITEL is operated by the Smithsonian 
Astrophysical Observatory (SAO) and was made possible by a grant from the 
Harvard University Milton Fund, a camera loan from the University of Virginia, 
and continued support of the SAO and UC Berkeley. The PAIRITEL project are 
further supported by NASA/{\it Swift} Guest Investigator grant NNG06GH50G and 
NNX08AN84G.  Some of the data presented herein were 
obtained at the W.~M.~Keck Observatory, which is operated as a scientific 
partnership among the California Institute of Technology, the University of
California and the NASA.
The Observatory was made possible by the generous financial support of
the W.~M.~Keck Foundation.  The authors wish to recognize and
acknowledge the very significant cultural role and reverence that the summit
of Mauna Kea has always had within the indigenous Hawaiian community.
We are most fortunate to have the opportunity to conduct observations from
this mountain.
}

{\it Facilities:} \facility{VLA}, \facility{HST (ACS)}, 
\facility{Swift (XRT)}, \facility{Keck:I (LRIS)},
\facility{PO:1.5m}, \facility{Hale (LFC)}, \facility{FLWO:PAIRITEL},
\facility{Spitzer (IRS)}, \facility{CARMA}



\clearpage
\begin{figure*}
\centerline{\plotone{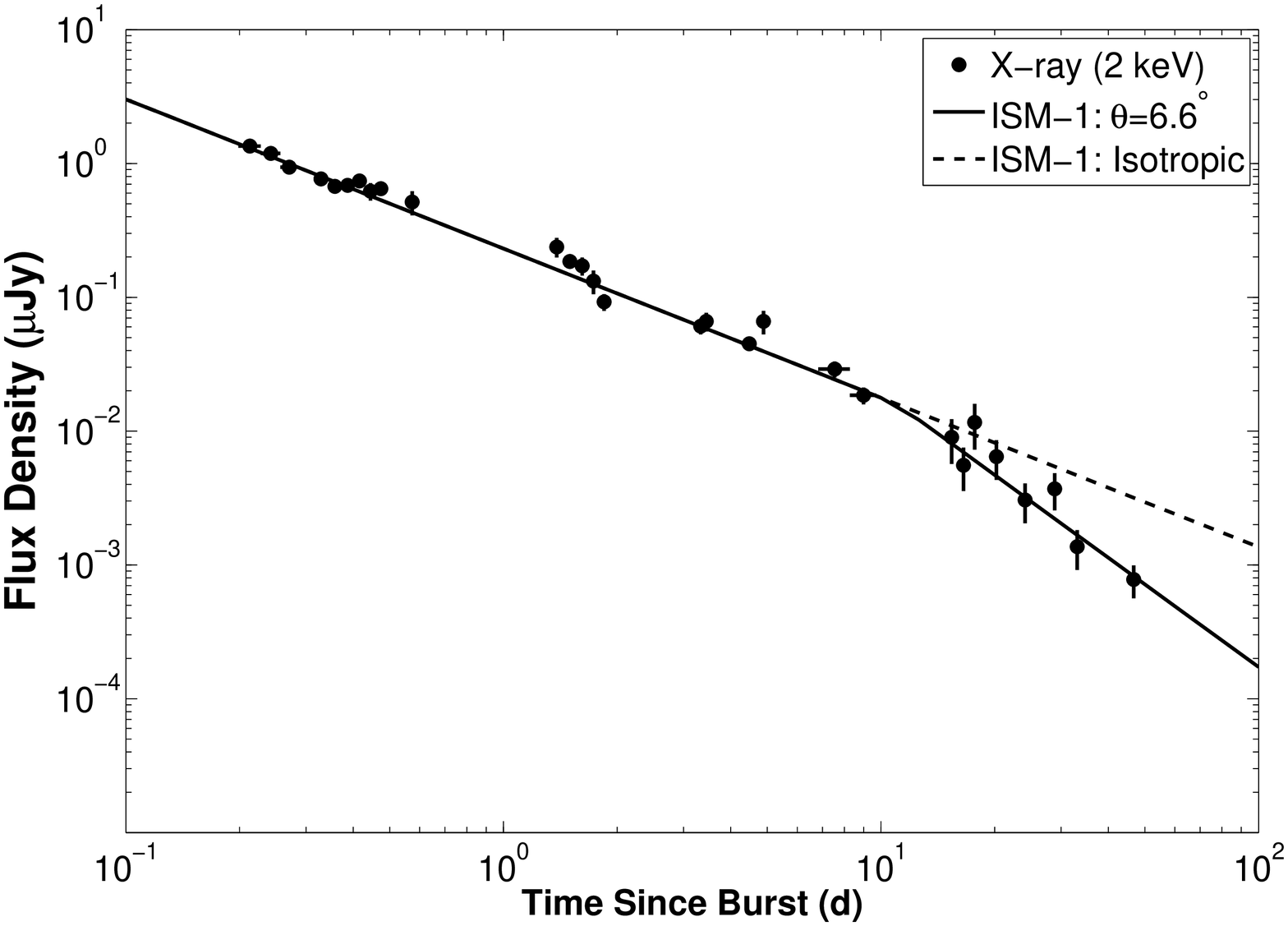}}
\centerline{\plotone{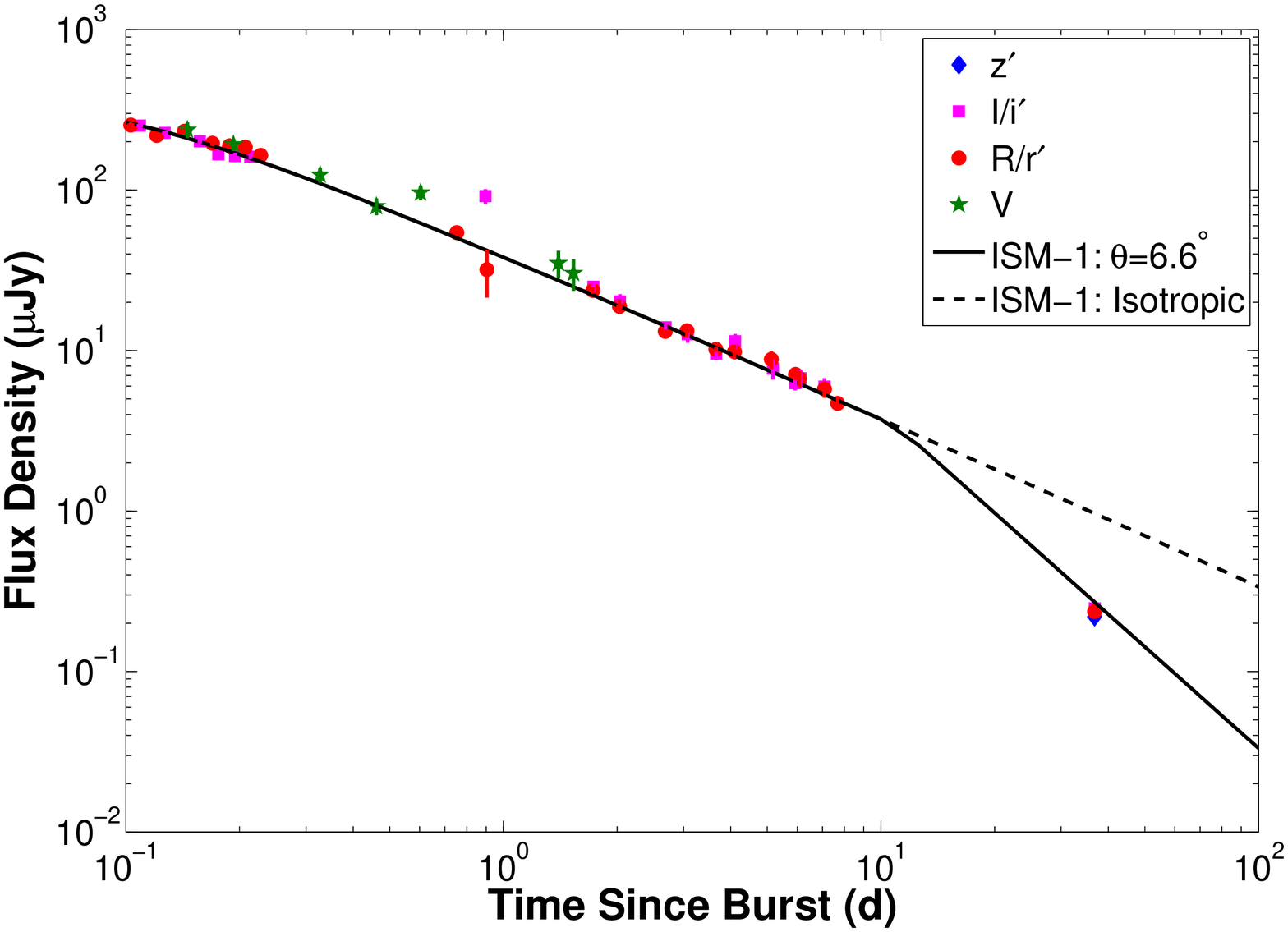}}
\end{figure*}
\clearpage

\clearpage
\begin{figure*}
\centerline{\plotone{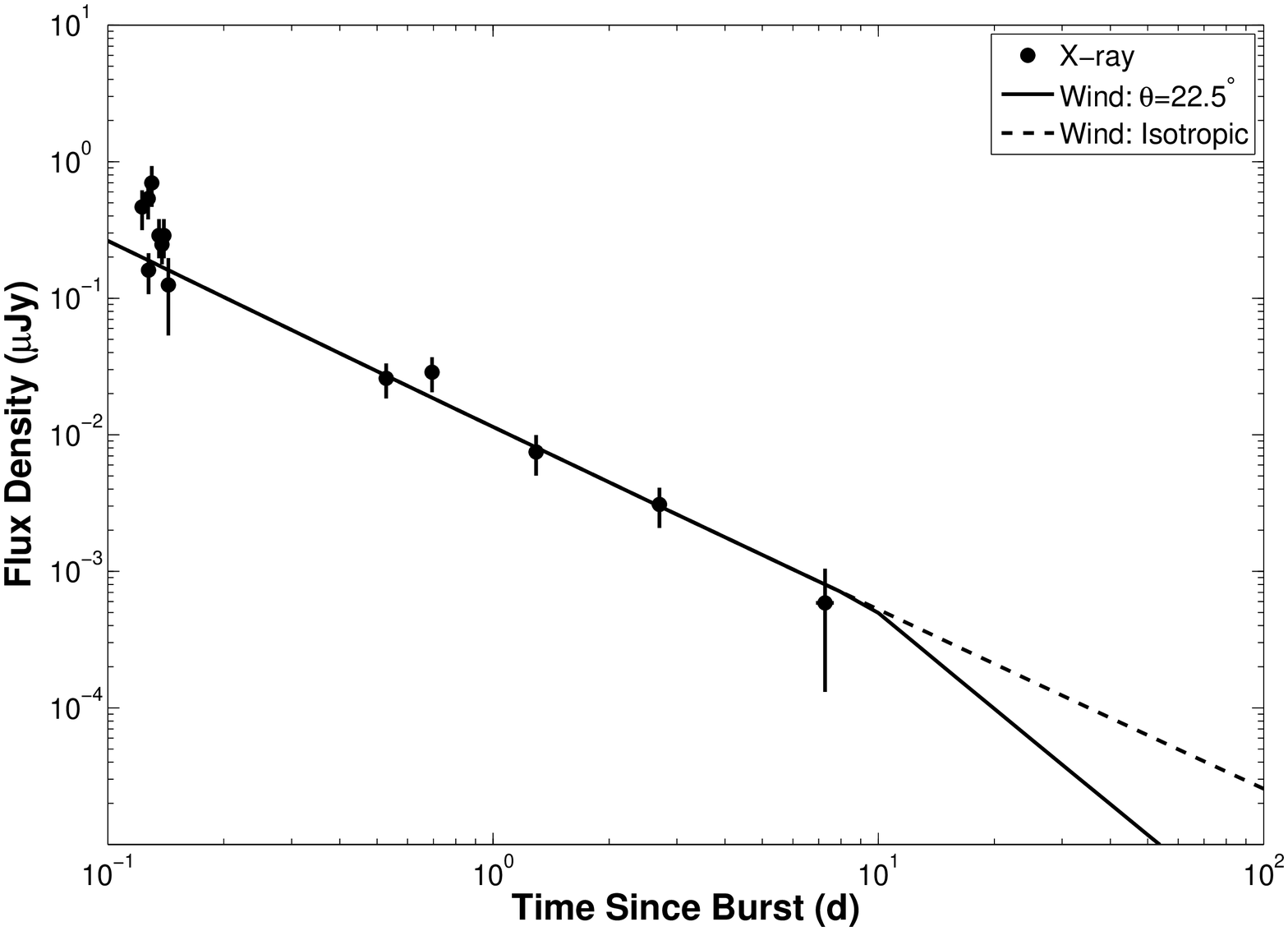}}
\centerline{\plotone{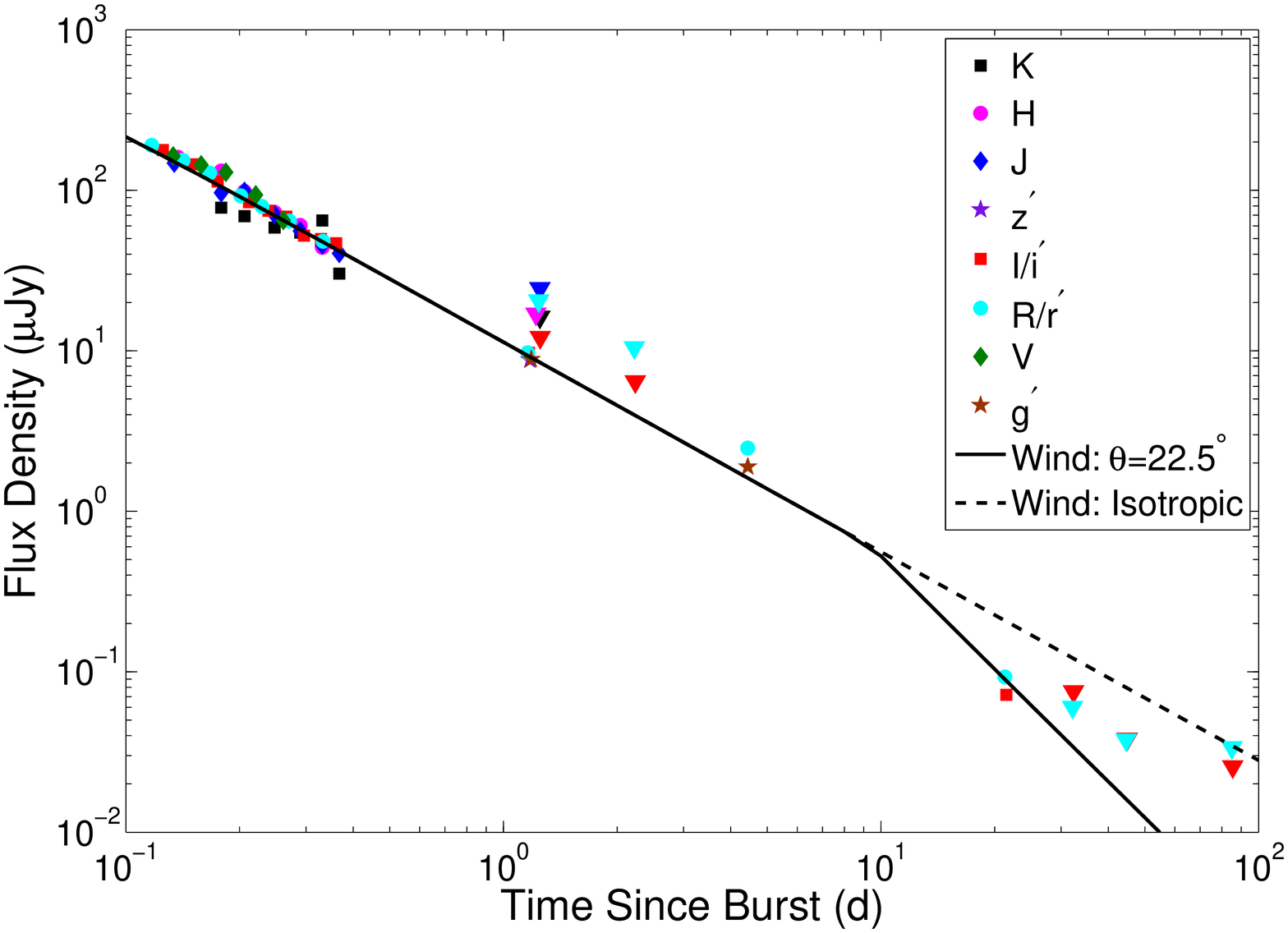}}
\end{figure*}
\clearpage

\clearpage
\begin{figure*}
\centerline{\plotone{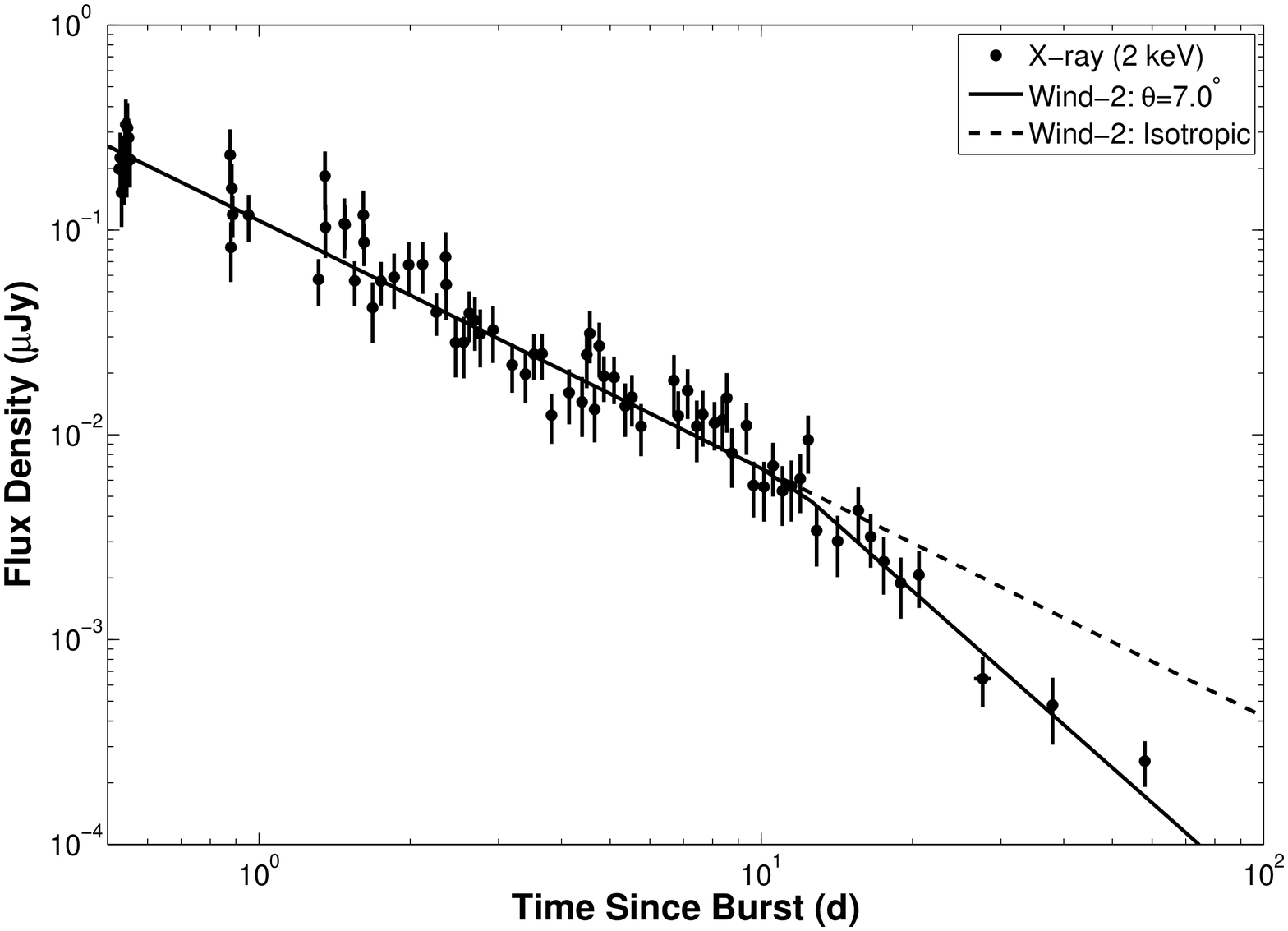}}
\centerline{\plotone{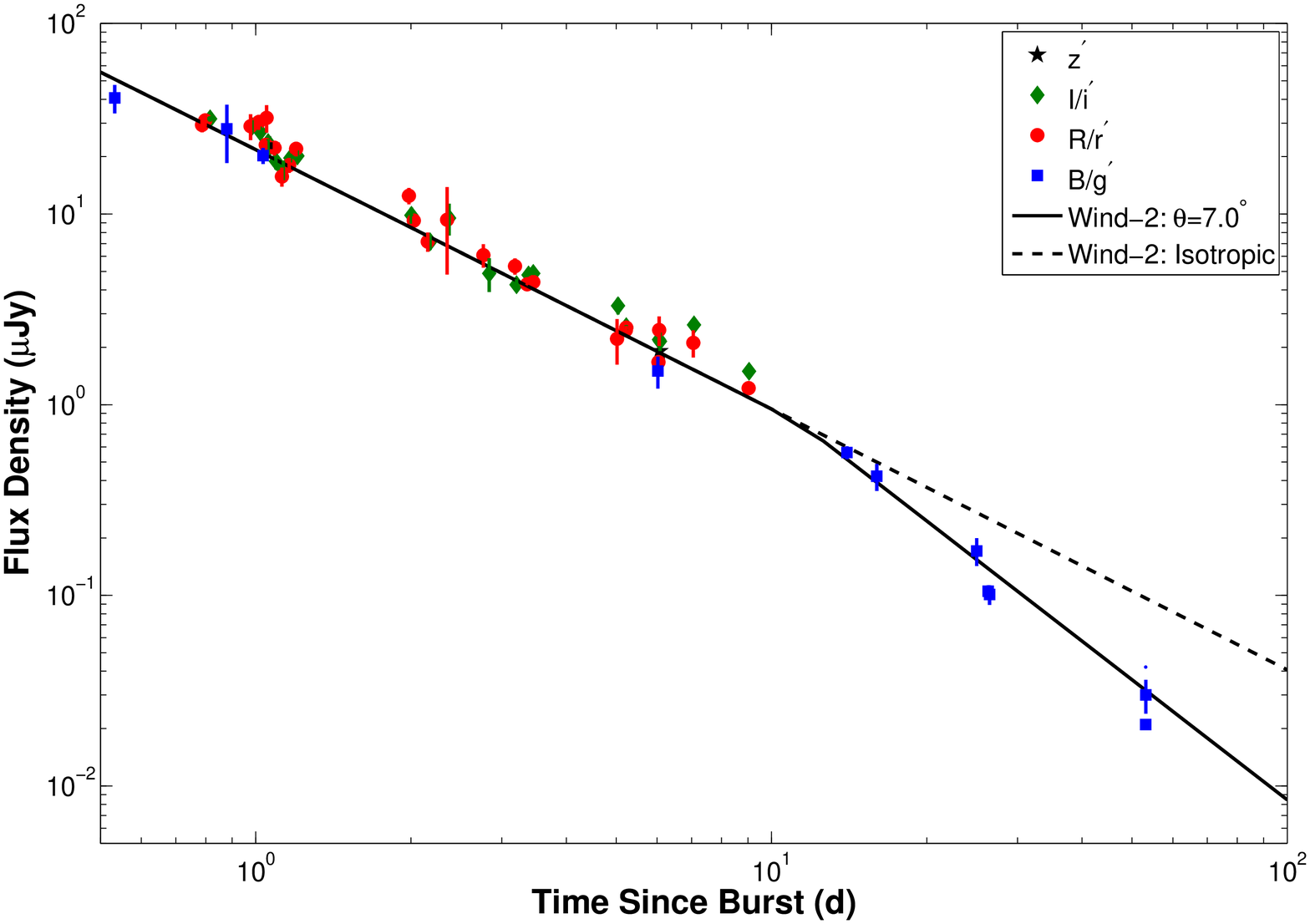}}
\end{figure*}
\clearpage

\clearpage
\input{tab7}

\end{document}